\documentclass[twocolumn,10pt]{IEEEtran}
\IEEEoverridecommandlockouts
\usepackage{amsmath}
\usepackage{amsfonts}
\usepackage{amssymb}
\usepackage{amsthm}
\usepackage{graphicx}
\usepackage{psfrag}
\usepackage{cite}
\usepackage{algorithm}
\usepackage{algorithmic}
\usepackage{color}
\usepackage{float}
\usepackage{epsfig,array,multicol,verbatim}
\usepackage{subfigure}
\usepackage{cases}
\usepackage {hhline}

\theoremstyle{plain} \newtheorem{theorem}{Theorem}
\theoremstyle{remark} \newtheorem{remark}{Remark}
\theoremstyle{plain} \newtheorem{lemma}{Lemma}
\theoremstyle{plain} \newtheorem{corollary}{Corollary}

\begin{document}
\title{Sustainable Throughput of Wireless LANs with Multi-Packet Reception Capability under Bounded Delay-Moment Requirements}
\author{Ying~Jun~(Angela)~Zhang,~\IEEEmembership{Member,~IEEE},  Soung~Chang~Liew,~\IEEEmembership{Senior Member,~IEEE}, and Da~Rui~Chen~\IEEEmembership{Student Member,~IEEE}
\thanks{This work was supported in part by the Competitive Ear-marked
Research Grant (Project Number 418707 and 414507) established under the
University Grant Committee of Hong Kong and the Direct Grant for Research (Project Numbers 2050439) established by The Chinese University of Hong Kong}
\thanks{The authors are with Dept. of Information Engineering, The Chinese University of Hong Kong. Email: \{yjzhang, soung, drchen\}@ie.cuhk.edu.hk}}

\maketitle

\begin{abstract}
With the rapid proliferation of broadband wireless services, it is of paramount importance to understand how fast data can be sent through a wireless local area network (WLAN). Thanks to a large body of research following the seminal work of Bianchi, WLAN throughput under saturated traffic condition has been well understood. By contrast, prior investigations on throughput performance under unsaturated traffic condition was largely based on phenomenological observations, which lead to a common misconception that WLAN can support a traffic load as high as saturation throughput, if not higher, under non-saturation condition. In this paper, we show through rigorous analysis that this misconception may result in unacceptable quality of service: mean packet delay and delay jitter may approach infinity  even when the traffic load is far below the saturation throughput. Hence, saturation throughput is not a sound measure of WLAN capacity under non-saturation condition. To bridge the gap, we define safe-bounded-mean-delay (SBMD) throughput and safe-bounded-delay-jitter (SBDJ) throughput that reflect the actual network capacity users can enjoy when they require finite mean delay and delay jitter, respectively.

 Our earlier work proved that in a WLAN with multi-packet reception (MPR) capability, saturation throughput scales super-linearly with the MPR capability of the network. This paper extends the investigation to the non-saturation case and shows that  super-linear scaling also holds for SBMD and SBDJ throughputs. Our results here complete the demonstration of MPR as a powerful capacity-enhancement technique for WLAN under both saturation and non-saturation conditions.

\end{abstract}


\section{Introduction}
\subsection{Motivation and Summary of Contributions}
Thanks to its simplicity, robustness and cost effectiveness, wireless local area networks (WLANs) based on IEEE 802.11 distributed control function (DCF) are playing a major role in next-generation home networks and hot spots. The recent explosion of broadband wireless services has stimulated considerable research interests in understanding how fast data can be sent through a WLAN. Since the seminal work of Bianchi in \cite{Bianchi:00}, there have been extensive efforts to characterize  throughput performance of WLANs under saturation condition, where stations always have packets to transmit \cite{Bianchi:00, Bianchi:05, Sakurai:07, Xiao:05}. By contrast, WLAN throughput in a non-saturated case is more complicated and less well understood, despite substantial recent efforts. Notably, there has been a common misconception that in a non-saturated case, WLAN can sustain a traffic load as high as saturation throughput while providing satisfactory quality of service (QoS). This misconception is backed up by recent work in \cite{Zhai:05, Duffy:05, Malone:07} which observed a throughput higher than saturation throughput before the network is saturated. In this paper, we argue that this misconception may lead to unbounded delay moments, which is not acceptable to most applications. In particular, our analysis reveals the facts that (i) the ``pre-saturation throughput peak" observed in \cite{Zhai:05, Duffy:05, Malone:07} only occurs under certain settings of backoff parameters; (ii) under many other circumstances, it is necessary to operate a WLAN \textit{far below} the saturation load to avoid unbounded mean packet delay and delay jitter.

The main objective of this paper is to answer the following important questions: (i) how much throughput can be sustained subject to finite mean delay and delay jitter in non-saturated WLANs; and (ii) how to maximize this throughput. To this end, we make the following contributions:

\begin{itemize}
  \item By investigating the operating point of DCF WLAN, we reveal the conditions under which the ``pre-saturation throughput peak" may occur. We show that this phenomenon is not true in general, except for certain backoff parameter settings.
  \item We propose a multiple-vacation queueing model to derive the explicit expressions for the probability distribution (in terms of transform) of packet delay. In contrast to existing work, our model captures the heavy-tail property of the probability distribution.

  \item Based on the heavy-tail delay distribution, we establish sufficient and necessary conditions for mean delay and delay jitter to be bounded. Notably, saturation throughput is no longer sufficient to characterize the network capacity in the unsaturated case, because mean delay and delay jitter may become unbounded prior to saturation. To bridge the gap, we define safe-bounded-mean-delay (SBMD) throughput and safe-bounded-delay-jitter (SBDJ) throughput as the maximum throughputs that can be safely sustained with finite mean delay and delay jitter, respectively.

  \item IEEE 802.11 DCF protocol adopts a backoff exponent $r=2$ in its exponential backoff (EB) mechanism. That is, when packets collide, the collision windows of the involved stations are doubled. In this paper, we show that oftentimes $r=2$ is not the optimal choice in terms of maximizing SBMD and SBDJ throughputs. In addition, compared with saturation throughput, SBMD and SBDJ throughputs are more sensitive to $r$. That is, one should be more careful in setting the right $r$ to avoid severe degradation of throughput. This paper shows how the optimal $r$ can be computed to maximize SBMD and SBDJ throughputs.

  \item In conventional WLANs, collision of packets occurs when more than one station transmits at the same time. With advanced PHY-layer signal processing techniques, it is possible for a receiver to detect multiple packets simultaneously without causing collisions in future WLANs. For example, in CDMA or multiple-antenna systems, multiple packets can be received simultaneously using multiuser detection (MUD) techniques \cite{Verdu:98}. This new collision model is referred to as multi-packet reception (MPR), as opposed to single-packet reception (SPR) in traditional WLANs. Our prior work in \cite{Zhang, Zheng:06ICC, Zheng:06WLN} shows that MPR can greatly enhance the capacity of WLANs: saturation throughput scales super-linearly with the MPR capability of the channel. In this paper, we extend the investigation to the non-saturation case and show that the maximum SBMD and SBDJ throughputs also scale super-linearly with the MPR capability. That is, SBMD and SBDJ throughputs divided by $M$ increases with $M$, where $M$ is the maximum number of packets that can be resolved simultaneously. Super-linear throughput scaling implies that the achievable throughput per unit cost increases with the MPR capability of the channel. This, together with our previous work in \cite{Zhang, Zheng:06ICC, Zheng:06WLN}, provides a strong incentive to deploy MPR in next generation wireless networks. Furthermore, it is shown that a large $M$ can decrease the sensitivity of SBMD and SBDJ throughputs to $r$. This provides another incentive to deploy MPR: the system is more robust against imprecise $r$ setting.
\end{itemize}

\subsection{Related Work}
Previous work on delay analysis can be divided into two main threads: medium-access delay of head-of-line (HOL) packets under saturation condition and queueing delay (also referred to as packet delay hereafter) under non-saturation condition. In saturated systems, mean medium-access delay is easily derived as the reciprocal of saturation throughput \cite{Bianchi:05, Xiao:05}. More recently, Sakurai and Vu derived moments and generating function of medium-access delay under saturation. It was found that the EB mechanism induces a heave-tailed delay distribution. Similar observation was also made by Yang and Yum in \cite{Yang:03} when binary EB is deployed. In this paper, we show that in unsaturated WLANs, packet delay distribution also exhibits a heavy-tail behavior. It is for precisely this reason that the sustainable throughput subject to finite mean delay and delay jitter may differ from the saturation throughput.

Packet delay under non-saturation condition has recently been analyzed in \cite{Zhai:05, Malone:07, Tickoo:04} using different techniques. Unfortunately, none of these analyses captures the heavy-tail characteristics of packet delay distribution. In this sense, the multiple-vacation queueing model proposed in this paper more accurately reflects the behavior of unsaturated WLANs.

Recent work in \cite{Zhai:05, Duffy:05, Malone:07} investigated the throughput performance under non-saturated operation. Notably, all the papers observed that a throughput higher than the saturation throughput occurs when network is not saturated. In this paper, we explain this phenomenon by analyzing the relationship between system operation point and saturation condition. In particular, it is shown that this phenomenon only exists in certain special cases. Moreover, even in these special cases, it is not wise to load the system with a traffic load higher than saturation throughput in practice, as such a traffic load cannot be safely sustained in the long run. On another front, \cite{Harsha:06} has analyzed system throughput in a more realistic scenario where VoIP and TCP traffics coexist in the network.

The work mentioned above are all based on SPR collision model. Previous work on delay analysis in unsaturated MPR networks has focused on pure slotted ALOHA systems \cite{Naware:05}. Apart from the underlying MAC protocol, a major difference between this paper and \cite{Naware:05} lies in the definition of MPR. In \cite{Naware:05}, average delay is characterized for a subclass of MPR channels, namely capture channels, where at most one user has a successful packet transmission when multiple packets contend for the channel at the same time. In \cite{Chan:04}, Chan and Berger analyzed the mean medium access delay for CSMA networks with MPR under saturation condition. Similar to \cite{Bianchi:05, Xiao:05}, the mean medium access delay is the reciprocal of saturation throughput in this case.

\subsection{Assumptions}
\begin{figure}[t]
\centering
\includegraphics[width=0.5\textwidth]{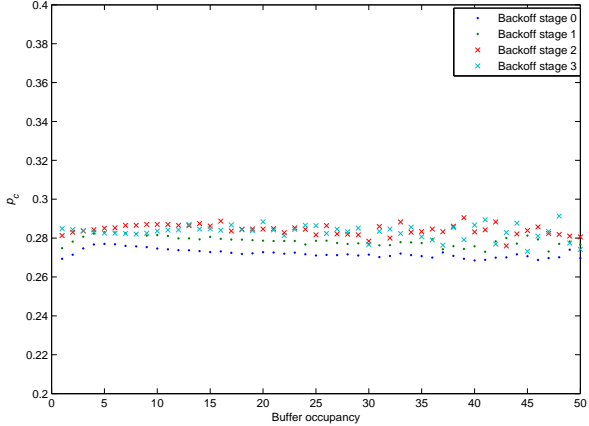}
\caption{Conditional collision probability $p_c$ for $N=10$ and $N\lambda=183$ packets per slot.} \label{validation}
\end{figure}

Similar to saturation analysis, most prior work on non-saturation analysis assumed that a station encounters a collision probability $p_c$ when it transmits, regardless of \textit{its own} backoff stage and buffer state \cite{Yang:03}-\cite{Tickoo:04}. Moreover, the transmission probability $\tau$ of an \textit{arbitrary} station in a \textit{generic} time slot is assumed to be constant and does not vary with the backoff stage and buffer state of a particular tagged station. This assumption is shown to be quite accurate when there are sufficient number of stations in the system. In this paper, we will adopt the same assumption and will discuss its impact on model accuracy in Section V-B. To validate this assumption, we simulate $p_c$ in Fig. \ref{validation} for the case of $N=10$ and $N\lambda=183$ packets per second. The figure shows that $p_c$ is largely independent of the backoff stage and buffer occupancy. In particular, the average variance of $p_c$ across buffer occupancy is about $7.9\times10^{-6}$ and that across backoff stage is about $3.73\times10^{-5}$. Despite this assumption, our model factors in the dependency of transmission probability of the tagged station on its backoff stage. It is this distinct feature of our model that enables us to capture the heavy-tail behavior of delay distribution.

Another assumption is that mobile stations keep monitoring the status of channel even when they have zero backlog in their queues. When a new packet arrives at an empty queue, it will start a backoff process at the beginning of the time slot that immediately follows. This assumption is slightly different from the standard IEEE 802.11 DCF. However, it makes our model sufficiently simple to obtain explicit expressions for quantities of interest, while still captures the dominant behaviors of non-saturated WLANs.

\subsection{Organization}
The remainder of the paper is organized as follows. In Section II, we briefly review the EB mechanism and the calculation of saturation throughput. System operation under non-saturation condition is discussed in Section III, where we establish the condition under which the system remains unsaturated. In Section IV, packet delay in EB-based WLAN is analyzed through an $M/G/1/V_m$ model. SBMD and SBDJ throughputs are defined in Section V, where we also show that the maximum SBMD and SBDJ throughputs scale super-linearly with the MPR capability of the network.  In Section VI, numerical results of two example systems are given to further illustrate our analysis. In Section VII, we extend our discussions to systems with a retry limit or a maximum contention window. Finally, the paper is concluded in Section VIII.

\section{Review of Exponential Backoff and Saturation Throughput}

We consider a fully-connected network with $N$ stations. The transmission of stations is coordinated by an EB mechanism. The EB mechanism adaptively tunes the transmission probability of a station according to the traffic intensity of the network. It works as follows. At each packet transmission, a station sets its backoff timer by randomly choosing an integer within the range $[0,W-1]$, where $W$ is the size of the contention window. The backoff timer is decreased by one following each time slot. The station transmits a packet from its buffer once the backoff timer reaches zero. At the first transmission attempt of a packet, $W$ is set to $W_0$, the minimum contention window. Each time the transmission is unsuccessful, $W$ is multiplied by a backoff factor $r$. That is, the contention window size $W_i=r^iW_0$ after $i$ successive transmission failures. For simplicity, we assume there is no retry limit in our system. However, our analysis can be easily extended to the case with a retry limit.

Thanks to advanced PHY-layer signal processing techniques, we assume that the channel has the capability to accommodate up to $M$ simultaneous packet transmissions. In other words, packets can be received correctly whenever the number of simultaneous transmissions is no larger than $M$. When more than $M$ stations contend for the channel at the same time, collision occurs and no packet can be decoded. By letting $M=1$, the system reduces to a traditional SPR WLAN. Our previous work in \cite{Zheng:06ICC, Zhang} has proposed one approach to implement MPR in IEEE 802.11 WLANs.

\begin{figure}
\centering
\includegraphics[width=0.45\textwidth]{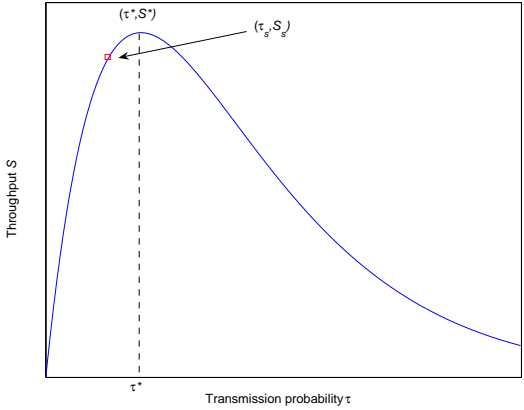}
\caption{$S-\tau$ curve} \label{fig:S-tau}
\end{figure}

Let $\tau$ denote the probability that a station transmits in a generic time slot. The probability of having $k$ stations transmit in a generic time slot is given by
\begin{eqnarray} \label{eq:X}
    \Pr\{X=k\}=\dbinom{N}{k}\tau^k(1-\tau)^{N-k}.
\end{eqnarray}
It is then straightforward to calculate the probabilities that a generic time slot is an idle slot, collision slot, and success slot in the following equations, where the superscript $G$ in
$P_{idle}^G$, $P_{coll}^G$ and $P_{succ}^G$ stands for ``generic''.
\begin{equation}\label{eq:P-idle}
      P_{idle}^G=\Pr\{X=0\}=(1-\tau)^N,
\end{equation}
\begin{equation}\label{eq:P-coll}
      P_{coll}^G=\Pr\{X\geq M+1\}=\sum^N_{k=M+1}\dbinom{N}{k}\tau^k(1-\tau)^{N-k},
\end{equation}
\begin{equation}\label{eq:P-succ}
      P_{succ}^G=\Pr\{X\leq M\}=\sum^M_{k=1}\dbinom{N}{k}\tau^k(1-\tau)^{N-k}.
\end{equation}
Given \eqref{eq:X}-\eqref{eq:P-succ}, WLAN throughput $S$, defined as the average number of information bits successfully transmitted per second, is given by \cite{Bianchi:00}
\begin{equation}\label{eq:S-sat}
    S=\frac{\sum^M_{k=1}k\Pr\{X=k\}PL}{P_{idle}^GT_{idle}+P_{coll}^GT_{coll}+P_{succ}^GT_{succ}},
\end{equation}
where $T_{succ}$, $T_{coll}$, and $T_{idle}$  denote the lengths of success, collision, and idle slots, respectively. $PL$ denotes the payload length of a packet. According to \eqref{eq:S-sat}, throughput $S$ can be plotted as a function of transmission probability $\tau$. One such example when $M=1$ and $T_{succ}=T_{coll}=T_{idle}$ is illustrated in Fig. \ref{fig:S-tau}.  The figure shows that the maximum throughput, $S^*$, occurs when $\tau$ is given by
\begin{equation}\label{eq:tau-opt}
    \tau^*=\arg\max\frac{PL\sum^M_{k=1}k\binom{N}{k}\tau^k(1-\tau)^{N-k}}{P_{idle}^GT_{idle}+P_{coll}^GT_{coll}+P_{succ}^GT_{succ}}.
\end{equation}

It is well known that under \textit{saturation} condition where stations are continuously backlogged, transmission probability $\tau$ is the root of a fixed-point system \cite{Bianchi:00}
\begin{equation}\label{eq:tau}
    \tau=\frac{2(1-rp_c)}{W_0(1-p_c)+1-rp_c},
\end{equation}
where $p_c$ is the probability that a station encounters collisions
when it transmits, which is given by
\begin{equation}\label{eq:pc}
    p_c=1-\sum^{M-1}_{k=0}\binom{N-1}{k}\tau^k(1-\tau)^{N-1-k}.
\end{equation}
To distinguish from the non-saturation case, we denote $\tau$ in the saturation case by $\tau_s$ in the rest of the paper. The above equations show that $\tau_s$ is a function of $r$. In Fig. \ref{fig:S-tau}, $\tau_s$ and the corresponding saturation throughput $S_s$ are plotted when $r=2$ and $W_0=16$.

Before leaving this section, we would like to reiterate that $r$ plays a major role in determining $\tau_s$ and $S_s$ \emph{under saturation condition}. That is, the saturation operating point of the system on the $S-\tau$ curve is determined by $r$. For the example in Fig. \ref{fig:S-tau}, $(\tau_s, S_s)$ lies to the left of the peak $(\tau^*, S^*)$ when $r=2$.

\section{System Operation Under Non-saturation Condition}
\subsection{Non-saturation throughput}
Under non-saturation, queues of stations are empty from time to time. Assume that packets arrive at each station according to a Poisson process at a rate of $\lambda$ packets per second. Let $\rho$ denote the probability that a queue is non-empty and $p_t$ denote the probability that a station with non-zero backlog transmits in a generic time slot. Then, the probability that a station transmits in a generic time slot, $\tau$, is linked to $p_t$ by
$\tau=\rho p_t+(1-\rho)\times 0=\rho p_t$.

We can derive the same expression as \eqref{eq:X} for $\Pr\{X=k\}$ in the non-saturation case as follows:
\begin{eqnarray} \label{eq:X-nonsaturation}
    &&\Pr\{X=k\}\nonumber\\
    &=&\sum_{j=k}^N \Pr\{k \text{ stations transmit}|j\text{ stations backlogged}\}\nonumber\\
    &&\times\Pr\{j\text{ stations backlogged}\}\nonumber\\
    &=&\sum_{j=k}^N\dbinom{N}{j}\rho^j(1-\rho)^{N-j}\dbinom{j}{k}p_t^k(1-p_t)^{j-k}\nonumber\\
    &=&\dbinom{N}{k}\tau^k(1-\tau)^{N-k}.
\end{eqnarray}
Likewise,  $P_{idle}^G$, $P_{coll}^G$, $P_{succ}^G$, and $p_c$ under non-saturation have the same forms as in \eqref{eq:P-idle}-\eqref{eq:P-succ} and \eqref{eq:pc}. The details are omitted due to the page limit.

Unlike saturation operation,  system throughput under non-saturation condition is always equal to the input traffic rate when the system is in a steady state\footnote{By steady state, we mean the Markov process associate with the queueing system observed by the stations is recurrent nonnull \cite{Kleinrock:75}. In other words, $\rho<1$, implying that the system is unsaturated and queues are empty from time to time.}. That is,
\begin{equation}\label{eq:S-unsat}
    N\lambda=\frac{\sum^M_{k=1}k\Pr\{X=k\}PL}{P_{idle}^GT_{idle}+P_{coll}^GT_{coll}+P_{succ}^GT_{succ}}.
\end{equation}
Solving \eqref{eq:S-unsat} yields the transmission probability $\tau$. A close observation of \eqref{eq:S-unsat} reveals the fact that unlike the saturation case, $\tau$ in the non-saturation case only depends on $N\lambda$ and is independent of $r$. Nonetheless, $r$ does have an effect on the saturation throughput $S_s$, which in turn determines whether a system can remain unsaturated under $N\lambda$.

\subsection{Operating point}
\begin{figure}
\centering
\includegraphics[width=0.5\textwidth]{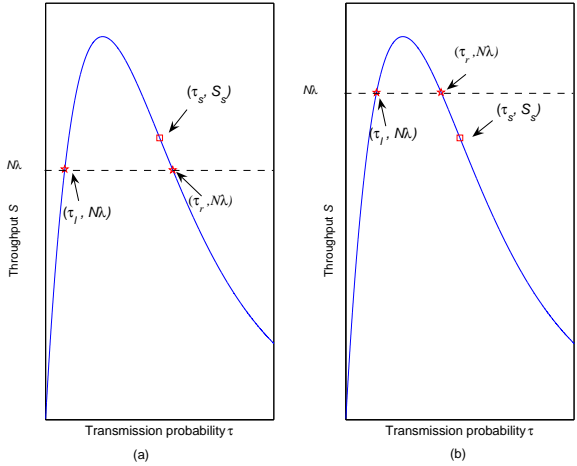}
\caption{Operating points: (a) $N\lambda<S_s$, $\tau_l<\tau_s<\tau_r$, (b) $N\lambda>S_s$, $\tau_l<\tau_r<\tau_s$} \label{fig:Nlambda}
\end{figure}

As shown in Fig. \ref{fig:Nlambda}, there could be two roots to \eqref{eq:S-unsat}, denoted by $\tau_l$ and $\tau_r$, respectively. However, Theorem \ref{theorem1} shows that only the root that is smaller than $\tau_s$ can be the system operating point under steady state.

\begin{theorem}\label{theorem1}
    A $\tau$ can be a steady-state operating point of WLANs if and only if $\tau<\tau_s$.
\end{theorem}

The proof is given in Appendix A. In the proof, we make use of some results that are derived in later sections. Readers are suggested to read the proof later after going through Section IV.

\begin{remark}
 Theorem \ref{theorem1} suggests that the network can  remain unsaturated (i.e., $\rho<1$), as long as $\tau<\tau_s$. Interestingly, when $\tau_s>\tau^*$, it is possible to have $N\lambda>S_s$ and $\tau_l<\tau_r<\tau_s$ (see Fig. \ref{fig:Nlambda}), implying the system can be unsaturated while having a throughput higher than $S_s$. This explains the ``pre-saturation throughput peak" observation in \cite{Zhai:05, Duffy:05, Malone:07}. It is worth noting that such phenomenon only occurs when the backoff parameters yield a $\tau_s$ larger than $\tau^*$ (If $\tau_s<\tau^*$, $N\lambda>S_s$ yields $\tau_s<\tau_l<\tau_r$, implying that the system cannot be stable in this case according to Theorem \ref{theorem1}.). In other words, it is not a general rule applicable to all WLAN settings.
\end{remark}

\section{Delay Analysis}
To understand how much throughput is sustainable subject to bounded delay moment requirements, we analyze packet delay of unsaturated WLAN in this section. In a fully connected network, each station sees the same environment. It is therefore sufficient to analyze the delay performance by investigating the queueing dynamic at one (arbitrarily chosen) tagged station.

Packet delay is composed of two parts: \emph{waiting time and medium-access delay}. In particular, waiting time denotes the time interval from the arrival of a packet to the instant when the packet becomes a HOL packet in the queue, and medium-access delay denotes the time period from the instant when the packet becomes a HOL packet to the instant at which the packet is successfully transmitted.

There is a strong temptation to model the system by an $M/G/1$ queue with medium-access delay being the service time. However, as we will elaborate in subsection IV-A and IV-B, the distribution of medium-access delay experienced by a packet depends on the buffer state seen by the packet upon its arrival. Consequently, the $M/G/1$ queueing model, which assumes service time is independent of buffer states, cannot be directly applied.

\subsection{Medium-access delay of packets that arrives at a non-empty queue}
\begin{figure*}[t]
\centering
\includegraphics[width=0.7\textwidth]{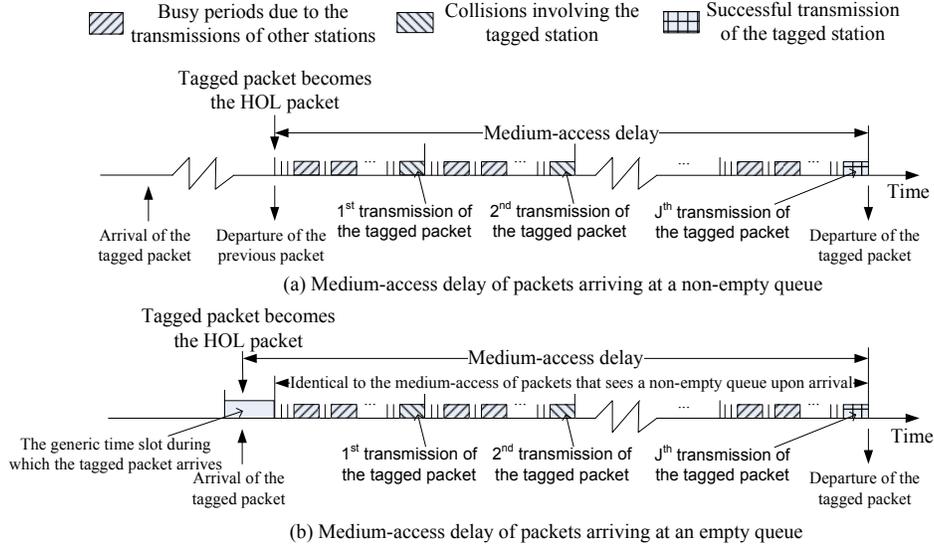}
\caption{Medium access delay.} \label{fig:delayIllustration}
\end{figure*}

A packet arriving at a non-empty queue becomes a HOL packet immediately after the preceding packet is successfully transmitted (assume the DIFS succeeding the transmission of the previous packet is included in the transmission time). Once it becomes a HOL packet, it starts a backoff process and attempts to access the channel
whenever the backoff counter reaches zero. As shown in Fig. \ref{fig:delayIllustration}a, there are three events that
contribute to the medium-access delay: backoff timer countdown, collisions involving the tagged station, and successful
transmissions of the tagged station. In particular, the backoff process consists of initial backoff and the backoff periods following unsuccessful transmissions of the tagged station. The probability that a packet is successfully transmitted on its $j^{th}$ transmission is given by
\begin{align}\label{eq:succ-j}
    \Pr\{R=j\}=p_c^{j-1}(1-p_c)\qquad \forall j\geq 1
\end{align}
where $R$ denotes the number of backoff periods that contributes to the medium-access delay.

The number of countdown slots in the backoff period between the $i-1^{th}$ and the $i^{th}$ transmission,
denoted by $B_i$, follows a discrete uniform distribution. That is,
\begin{equation}\label{eq:B}
    \Pr\{B_i=k\}=\frac{1}{r^{i-1}W_0} \forall k\in [0,r^{i-1}W_0-1],
\end{equation}
with the corresponding $Z$ transform being
\begin{equation}\label{eq:B-Z}
    \hat{B}_i(Z)=\sum^{r^{i-1}W_0-1}_{k=0}\Pr\{B_i=k\}z^k=\frac{1-z^{r^{i-1}W_0}}{r^{i-1}W_0(1-z)}
\end{equation}
Meanwhile, each countdown slot can be either idle or occupied by collisions and successful transmissions \emph{not} involving the tagged station  with the following probabilities: (where superscript $B$ stands for time slots in the
``backoff'' process of the tagged station)
\begin{equation}\label{eq:P-idle-B}
    P_{idle}^B=(1-\tau)^{N-1},
\end{equation}
\begin{equation}\label{eq:P-coll-B}
    P_{coll}^B=1-\sum^M_{k=0}\binom{N-1}{k}\tau^k(1-\tau)^{N-1-k},
\end{equation}
and
\begin{equation}\label{eq:P-succ-B}
    P_{succ}^B=\sum^M_{k=1}\binom{N-1}{k}\tau^k(1-\tau)^{N-1-k}.
\end{equation}
Let $L$ be a random variable denoting the length of a countdown
slot. The Laplace transform of $L$ is given by
\begin{eqnarray}\label{eq:L}
    L^*(s)&=&\mathrm{E}[e^{-sL}]\\
    &=&e^{-sT_{idle}}P_{idle}^B+e^{-sT_{coll}}P_{coll}^B+e^{-sT_{succ}}P_{succ}^B.\nonumber
\end{eqnarray}
The duration of the backoff period between the
$i-1^{th}$ and the $i^{th}$ transmission, denoted by $C_i$, can now be calculated as the sum of
the lengths of $B_i$ countdown time slots. That is,
\begin{equation}\label{eq:Ci}
    C_i=\sum^{B_i}_{i=1}L_m,
\end{equation}
where $L_m$ denotes the length of the $m^{th}$ countdown time slot, which are
identically and independently distributed (i.i.d.). The Laplace
transform of $C_i$ is calculated as
\begin{eqnarray}\label{eq:Ci-trans}
        C_i^*(s)&=&\mathrm{E}\left[e^{-sC_i}\right]\nonumber\\
        &=&\sum^{r^{i-1}W_0-1}_{k=0}\prod^k_{m=1}\mathrm{E}\left[\exp(-sL_m)\right]\Pr\{B_i=k\}\nonumber\\
        &=&\sum^{r^{i-1}W_0-1}_{k=0}\left(L^*(s)\right)^k\Pr\{B_i=k\}\nonumber\\
        &=&\hat{B}_i(L^*(s)).
\end{eqnarray}
We are now ready to derive the medium-access delay of packets that
arrive at a non-empty queue, denoted by $X_{ne}$, as follows:
\begin{equation}\label{eq:Xne}
    X_{ne}=\sum^j_{i=1}C_i+(j-1)T_{coll}+T_{succ} \quad \text{if }R=j,
\end{equation}
with the Laplace transform being
\begin{eqnarray}\label{eq:Xne-trans}
      X_{ne}^*(s)&=&\sum^\infty_{j=1}\prod^j_{i=1}\mathrm{E}\left[e^{-sC_i}\right]e^{ -s\left(\left(j-1\right)T_{coll}+T_{succ}
        \right)}\Pr\{R=j\} \nonumber\\
      &=&\sum^\infty_{j=1}
      e^{-s\left(\left(j-1\right)T_{coll}+T_{succ}\right)}\Pr\{R=j\}\prod^j_{i=1}\hat{B}_i\left(L^*(s)\right)\nonumber\\
\end{eqnarray}
where the first equality is due to the fact that $C_i$'s are
independent for different $i$'s.
\begin{theorem}\label{convergence}
The $n^{th}$ moment of $X_{ne}$ is finite if and only if $p_c<\frac{1}{r^n}$.
\end{theorem}
\noindent\emph{Proof:} To prove Theorem \ref{convergence}, we first investigate the form of $\mathrm{E}[C_i^n]$.
\begin{eqnarray}\label{C-moment}
&&\mathrm{E}[C_i^n]=(-1)^n\frac{d^n\hat{B}_i(L^*(s))}{ds^n}\bigg|_{s=0}\\
&=&(-1)^n \frac{d^n\hat{B}_i(L^*(s))}{dL^*(s)^n}\bigg|_{s=0}\bigg(\frac{dL^*(s)}{ds}\bigg)^n\bigg|_{s=0}\nonumber\\&&+\text{other terms with lower order derivatives of}~ \hat{B}_i\nonumber\\
&=&(T_{idle}P_{idle}^B +T_{coll}P_{coll}^B+T_{succ}P_{succ}^B)^n\frac{r^{(i-1)n}W_0^n}{n+1}\nonumber\\
&&+\text{other terms with power of $r$ lower than $(i-1)n$}.\nonumber
\end{eqnarray}
From \eqref{eq:Xne}, we have
\begin{eqnarray}\label{Xne-moment}
&&\mathrm{E}[X_{ne}^n]\nonumber\\
&=&(1-p_c)\sum^\infty_{j=1}p_c^{j-1}\mathrm{E}\big[\big(\sum^j_{i=1}C_i+(j-1)T_{coll}+T_{succ}\big)^n\big]\nonumber\\
&=&(1-p_c)\sum^\infty_{j=1}p_c^{j-1}\bigg(\mathrm{E}[C_j^n]+\text{other terms with} \nonumber\\&&\text{power of $r$ lower than}~(j-1)n\bigg)
\end{eqnarray}
Combining \eqref{C-moment} and \eqref{Xne-moment}, it is easy to see that $\mathrm{E}[X_{ne}^n]$ contains the term
\begin{eqnarray}\label{term}
&&(T_{idle}P_{idle}^B+T_{coll}P_{coll}^B+T_{succ}P_{succ}^B)^n\frac{W_0^n(1-p_c)}{n+1}\nonumber\\
&\times&\sum^\infty_{j=1}p_c^{j-1}r^{(j-1)n},
\end{eqnarray}
which converges if and only if $p_c<\frac{1}{r^n}$. All other terms in $\mathrm{E}[X_{ne}^n]$ involve lower powers of $r$. As $r>1$ by definition, it is straightforward that \eqref{Xne-moment} converge to a finite value as long as \eqref{term} converges. This completes the proof.
$\hfill\blacksquare$

As special cases, the first three moments of $X_{ne}$ converges to the forms in \eqref{eq:EXneConv}-\eqref{eq:EXne3Conv} when $p_c$ is smaller than $1/r$, $1/r^2$, and $1/r^3$, respectively. (\eqref{eq:EXne2Conv} is shown at top of next page.)
\begin{equation}\label{eq:EXneConv}
    \mathrm{E}[X_{ne}]=A_1\frac{W_0(1-p_c)-(1-rp_c)}{2(1-p_c)(1-rp_c)}+T_{coll}\frac{p_c}{1-p_c}+T_{succ}.
\end{equation}

\begin{table*}[ht]
\begin{eqnarray}\label{eq:EXne2Conv}
      &&E\left[X_{ne}^2\right]=A_1^2\begin{pmatrix}\frac{W_0^2}{12(1-r^2p_c)}-\frac{W_0}{(1-rp_c)}+\frac{5}{12(1-p_c)}+\frac{W_0^2(1+rp_c)}{4(1-rp_c)(1-r^2p_c)} -\frac{W_0^2(1-rp_c^2)}{2(1-p_c)(1-rp_c)^2}
        +\frac{1+p_c}{4(1-p_c)^2}\end{pmatrix} \nonumber\\
      &+&\begin{pmatrix}A_2\frac{W_0(1-p_c)-(1-rp_c)}{2(1-rp_c)(1-p_c)}+T_{coll}^2\frac{p_c(1+p_c)}{(1-p_c)^2}+T_{succ}^2
        +2A_1T_{coll}p_c\left( \frac{W_0(1+r-2rp_c)}{2(1-p_c)(1-rp_c)^2}
        -\frac{1}{(1-p_c)^2}\right)\end{pmatrix} \nonumber\\
      &+&\begin{pmatrix}A_1T_{succ}\frac{W_0(1-p_c)-(1-rp_c)}{(1-rp_c)(1-p_c)}
        +2T_{coll}T_{succ}\frac{p_c}{(1-p_c)}\end{pmatrix}.
\end{eqnarray}
\end{table*}
\begin{equation}\label{eq:EXne3Conv}
    E\left[X_{ne}^3\right]=\theta_1+3\theta_2+\theta_3,
\end{equation}
where $\theta_1$, $\theta_2$, and $\theta_3$ are given in Appendix B. Moreover,
\begin{eqnarray}
A_1&=&T_{idle}P_{idle}^B+T_{coll}P_{coll}^B+T_{succ}P_{succ}^B,\nonumber\\ A_2&=&T_{idle}^2P_{idle}^B+T_{coll}^2P_{coll}^B+T_{succ}^2P_{succ}^B,\nonumber\\
\text{and } A_3&=&T_{idle}^3P_{idle}^B+T_{coll}^3P_{coll}^B+T_{succ}^3P_{succ}^B.\nonumber
\end{eqnarray}

\begin{remark}\label{remark:condition}
In EB schemes where $r>1$, $p_c<1/r^{n_1}$ is a tighter condition
than $p_c<1/r^{n_2}$ if $n_1>n_2$. In other words, the convergence
of $\mathrm{E}[X_{ne}^{n_1}]$ implies the convergence of $\mathrm{E}[X_{ne}^{n_2}]$
for $n_1>n_2$, but not the reverse.
\end{remark}

\noindent\textbf{Definition:} A random variable is said to have a heavy tail distribution if it does not have all its power moments finite \cite{heavytail}.

\begin{remark}
Theorem \ref{convergence} shows that $X_{ne}$ has a heavy tail distribution. This is because for any given $r>1$, there always exists an $N$ such that $p_c<1/r^n$ for all $n\leq N$, but $p_c\geq1/r^n$ for all $n>N$. This implies that $X_{ne}$ does not have all its power moments finite. Similar observation was made in \cite{Sakurai:07} for medium-access delay under saturated operation. As we will show shortly, such feature of $X_{ne}$ leads to a heavy tail distribution of the packet delay of non-saturated WLANs.
\end{remark}

\subsection{Medium-access delay of packets that arrive at an empty queue}
Packets that arrive at empty queues undergo a different medium
access delay than that derived in the last subsection. As shown in Fig. \ref{fig:delayIllustration}b, a packet that arrives at an empty
queue becomes a HOL packet immediately after its arrival. The
arrival may occur in the midst of an idle time slot or a time slot
that is occupied by collisions or successful transmissions of other
stations. According to
the protocol, the packet must wait until the end of
this time slot before it has a chance to be selected for transmission. Once the time slot
ends, the backoff process starts and the packet will be transmitted
once the backoff timer counts down to zero. When $N$ is
relatively large, the channel states of adjacent time slots are
effectively independent of each other. As a result, the backoff
process, once started, is stochastically identical to the one
described in Section IV-A. In other words, as illustrated in Fig. \ref{fig:delayIllustration}b, the medium-access delay of packets
arriving at an empty queue, denoted by $X_e$, consists of two parts:
a time period that is statistically identical to $X_{ne}$ and an
interval between the arrival time and the time instant at which the backoff process starts.
That is,
\begin{equation}\label{eq:egtne}
    X_e= Y+X_{ne},
\end{equation}
where $Y$ is the random variable denoting the remaining period of the time slot during which the packet arrives.

In the next subsection, we will show that such a system behavior is naturally modeled by an $M/G/1/V_m$ ($M/G/1$ with multiple vacations) queueing
model.

\subsection{$M/G/1/V_m$ queueing model}
As discussed in the two subsections above, as long as the tagged station is continuously backlogged, all packets
 experience a service time of $X_{ne}$. Once the station becomes idle,
the channel will be occupied by an idle time slot or a busy time
slot due to the transmissions of \emph{other} stations. If the
tagged station is still empty at the end of this time slot, the
channel will be occupied by another time slot that is i.i.d.
to the previous one. Otherwise, the tagged station will start a backoff process at the end of the slot and its HOL packet will have a chance to access the channel.

The aforementioned behavior is well modeled by an
$M/G/1/V_m$ queue \cite{Doshi:86} where the server takes a vacation
every time the system becomes empty. In our case, the time slots that are occupied by other stations or idle slots when the tagged station is empty can be modeled as vacations to the tagged station. It is easy to show that a vacation
period has the same distribution as in \eqref{eq:L}. Then, $Y$ in eqn. \eqref{eq:egtne} is the
forward recurrence time of the vacation and has a Laplace
transform of
\begin{eqnarray}\label{eq:Y}
&& Y^*(s) = \frac{1-L^*(s)}{s\mathrm{E}[L]} \\
 &=&  \frac{1-\left(e^{-sT_{idle}}P^B_{idle}+e^{-sT_{coll}}P^B_{coll}+e^{-sT_{succ}}P^B_{succ}\right)}
        {s\left(T_{idle}P^B_{idle}+T_{coll}P^B_{coll}+T_{succ}P^B_{succ}\right)}.\nonumber
\end{eqnarray}
It can be easily shown that $\mathrm{E}[Y]=\tfrac{A_2}{2A_1}$ and
$\mathrm{E}[Y^2]=\tfrac{A_3}{3A_1}$.

We are now ready to calculate packet delay of WLAN, denoted by $D$, as the system time of an $M/G/1/V_m$ queue with arrival rate $\lambda$, service time $X_{ne}$, vacation time $L$ with forward recurrence time $Y$. As is standard in $M/G/1/V_m$ queue analysis, packet delay $D$ follows a distribution of \cite{Doshi:86}
\begin{equation}\label{eq:Q-s}
      D^*(s) =(1-\tilde{\rho})X_{ne}^*(s)\frac{sY^*(s)}{\lambda X^*_{ne}(s)-\lambda +s},
\end{equation}
where $\tilde{\rho}=\lambda\mathrm E[X_{ne}]$ can be regarded as the server utilization of an $M/G/1$ queue without vacation but with a service time $X_{ne}$. The queue length, $Q$, therefore follows the following distribution: \cite{Kleinrock:75}
\begin{eqnarray}
\hat{Q}(z)&=&D^*(\lambda-\lambda z)\nonumber\\
&=&(1-\tilde{\rho})X_{ne}^*(\lambda - \lambda z)\frac{(1-z)Y^*(\lambda - \lambda z)}{X_{ne}^*(\lambda - \lambda z)-z}.
\end{eqnarray}

It is now straightforward that mean delay $\mathrm{E}[D]$ and delay jitter $\mathrm{Var}[D]$ are given by

\begin{eqnarray}\label{eq:ED}
    \mathrm{E}[D]&=&-\frac{\mathrm{d}D^*(s)}{\mathrm{d}s}\bigg|_{s=0}\nonumber\\
    &=&\mathrm{E}[X_{ne}]+\mathrm{E}[Y]+\frac{\lambda\mathrm{E}[X^2_{ne}]}{2(1-\tilde{\rho})},
\end{eqnarray}
and
\begin{eqnarray}\label{eq:VarD}
      &&\mathrm{VAR}[D]= \mathrm{E}\left[D^2\right]-\left(\mathrm{E}[D]\right)^2\\
        &=&\frac{\mathrm{d^2}D^*(s)}{\mathrm{d}s^2}\bigg|_{s=0}-\left(\mathrm{E}[D]\right)^2\nonumber\\
        &=&\mathrm{VAR}[X_{ne}]+\mathrm{VAR}[Y]
        +\frac{\lambda^2\left(E\left[X_{ne}^2\right]\right)^2}{4(1-\tilde{\rho})^2}+\frac{\lambda
        \mathrm{E}[X^3_{ne}]}{3(1-\tilde{\rho})}.\nonumber
\end{eqnarray}
The utilization of the server in the $M/G/1/V_m$ system, denoted by
$\rho$, is given by
\begin{eqnarray}\label{eq:rho}
      \rho&=& 1-\Pr\{Q=0\}=1-\hat{Q}(0) \\
        &=&1-(1-\tilde{\rho})\nonumber\\
        &&\times\frac{1-\left(e^{-\lambda T_{idle}}P^B_{idle}+
        e^{-\lambda T_{coll}}P^B_{coll}+e^{-\lambda T_{succ}}P^B_{succ}\right)}
        {\lambda\left(T_{idle}P^B_{idle}+T_{coll}P^B_{coll}+T_{succ}P^B_{succ}\right)}.\nonumber
\end{eqnarray}
According to queueing theory, a Markov chain associated with a queue
can reach a steady state if and only if $\rho<1$
\cite{Kleinrock:75}. It is not difficult to see from \eqref{eq:rho}
that the following statements are equivalent:
\begin{itemize}
  \item The network is unsaturated.
  \item $\tilde{\rho}<1$.
  \item $\rho<1$.
  \item $\rho>\tilde{\rho}$.
\end{itemize}
Moreover, when the network is saturated, $\rho=\tilde{\rho}=1$.

\begin{theorem}\label{theorem:finitedelay}
Mean packet delay $\mathrm{E}[D]$ in \eqref{eq:ED} is finite if and only if $\tilde{\rho}<1$ and
$p_c<1/r^2$. Likewise, delay jitter $\mathrm{VAR}[D]$ in \eqref{eq:VarD} is finite if and only
if $\tilde{\rho}<1$ and $p_c<1/r^3$.
\end{theorem}
\noindent\emph{Proof:}
The proof is straightforward from
the conditions for convergence of the first three moments of
$X_{ne}$ and the fact that $\tilde{\rho}=\lambda \mathrm{E}[X_{ne}]<1$ implies $\mathrm{E}[X_{ne}]$
is finite.
$\hfill\blacksquare$

\begin{corollary}\label{corollary:finitedelay}
When the number of stations $N$ is large, $\mathrm{E}[D]$ is finite if and only if $p_c<1/r^2$, and $\mathrm{Var}[D]$ is finite if and only if $p_c<1/r^3$.
\end{corollary}
\noindent\emph{Proof:}
Considering \eqref{eq:EXneConv} and the fact that $\tilde{\rho}=\lambda\mathrm{E}[X_{ne}]$, the non-saturation condition $\tilde{\rho}<1$ can be equivalently written as
\begin{equation}\label{eq:rho<1}
    p_cr+\frac{\lambda A_1W_0(1-p_c)}{2(1-p_c)-2\lambda T_{succ}(1-p_c)-2\lambda T_{coll}p_c+\lambda
    A_1}<1.
\end{equation}
When $N$ is large, $\lambda\rightarrow 0$ because system throughput $N\lambda$ is finite. As a result, the second term in the left hand side of \eqref{eq:rho<1} converges to zero. In this case, the inequality $\tilde{\rho}<1$ reduces to $p_c<1/r$, which is automatically satisfied when $p_c<1/r^2$ (or $p_c<1/r^3$). Therefore, for large $N$, $p_c<1/r^2$ is the sufficient and necessary condition for finite mean delay $\mathrm{E}[D]$, while $p_c<1/r^3$ is the sufficient and necessary condition for finite delay jitter $\mathrm{\mathrm{VAR}}[D]$.
$\hfill\blacksquare$

\begin{remark}\label{remark4}
It is obvious from Remark \ref{remark:condition} and Theorem \ref{theorem:finitedelay} that mean delay $\mathrm{E}[D]$ and delay jitter $\mathrm{VAR}[D]$ can be infinite even if the system is unsaturated.
\end{remark}
Readers are now ready to read the proof of Theorem \ref{theorem1}, which is given in Appendix A.

\section{SBMD and SBDJ Throughput}
It is generally accepted that a traffic load is sustainable as long
as it is lower than saturation throughput. However, Remark \ref{remark4} reveals
the fact that packets may suffer from very large mean delay or delay
jitter even for traffic loads lower than saturation throughput. In many
applications, it is crucial to guarantee bounded mean delay or delay
jitter. To bridge the gap, we will define in this section SBMD and
SBDJ throughputs, which are the highest throughputs that can be
sustained with bounded mean delay and delay jitter, respectively.

\subsection{Boundary-bounded-mean-delay (BBMD) and boundary-bounded-delay-jitter (BBDJ) throughput}
For large $N$, $p_c<1/r^2$ and $p_c<1/r^3$ are sufficient and
necessary conditions for bounded mean delay and bounded delay jitter
respectively. By observing the boundary cases where $p_c=1/r^2$ and
$p_c=1/r^3$, we can get the \emph{highest} possible transmission
probabilities that do not cause unbounded mean delay and bounded
delay jitter, respectively. Denote such transmission probabilities
by $\tau_{BBMD}$ and $\tau_{BBDJ}$, respectively, and the
corresponding throughput by $S_{BBMD}$ and $S_{BBDJ}$. It is obvious
from \eqref{eq:pc} that $p_c$ is an increasing function of $\tau$.
Hence, $\tau_{BBDJ}<\tau_{BBMD}<\tau_{s}$.

\begin{figure}
\centering
\includegraphics[width=0.5\textwidth]{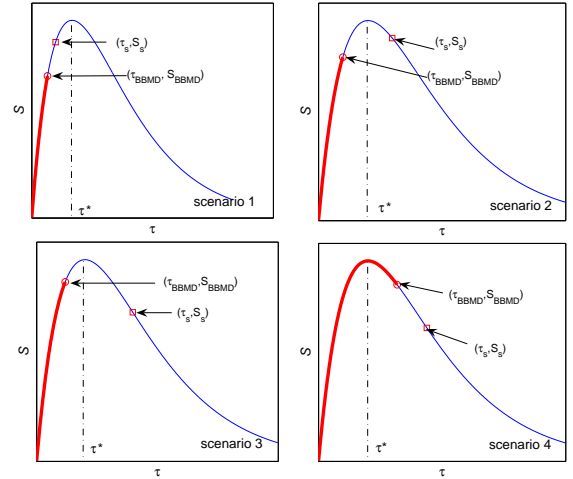}
\caption{Scenario 1: $\tau_{BBMD}<\tau_s<\tau^*$, $S_{BBMD}<S_s$;
scenario 2: $\tau_{BBMD}<\tau^*<\tau_s$, $S_{BBMD}<S_s$; Scenario 3:
$\tau_{BBMD}<\tau^*<\tau_s$, $S_{BBMD}>S_s$; scenario 4:
$\tau^*<\tau_{BBMD}<\tau_s$, $S_{BBMD}>S_s$.}
 \label{fig:4_scenarios}
\end{figure}

As discussed in Section II, depending on $r$ and other system parameters, $\tau_s$ can be
smaller than, equal to, or larger than $\tau^*$. The relationship
between $S_s$, $S_{BBMD}$, and $S_{BBDJ}$ highly depends on the
position of $\tau_s$. To show this, we illustrate four different
scenarios in Fig. \ref{fig:4_scenarios}: $\tau_s\leq\tau^*$ in
scenario 1 and $\tau_s>\tau^*$ in the other 3 scenarios. For simple
illustration, we focus on $S_{BBMD}$ only. However, the following
conclusions can be easily extended to $S_{BBDJ}$ by replacing the
inequality $p_c<1/r^2$ with $p_c<1/r^3$.

In Fig. \ref{fig:4_scenarios}, the thickened parts of the curves
denote the region in which $p_c<1/r^2$. Operating regions beyond the
thickened part in each of the scenarios is not viable if bounded
mean delay is to be achieved. In scenarios 1, 2, and 3, mean packet
delay becomes unbounded when the input traffic load $N\lambda$ is
higher than $S_{BBMD}$. In particular in scenarios 1 and 2 where
$S_{BBMD}<S_s$, it is necessary to load the system below the
saturation point by a sufficient margin to avoid excessively long packet delay.

In scenarios 3 and 4, $S_{BBMD}>S_s$. In these cases, it is
\emph{theoretically} possible to operate the system at a higher
throughput than the saturation throughput $S_s$ while achieving a
bounded mean delay. More interestingly, in scenario 4, it is even
possible to load the system at the maximum throughput $S^*$ while
having a finite mean delay \emph{in theory}, as long as the system
is operated within the thickened region of the curve. However, as we
will argue in the next subsection, it is not safe to load the system
with an offered load higher than $S_s$.

So far, we have discussed the large $N$ case where $p_c<1/r^2$ (resp. $p_c<1/r^3$) is a stricter condition than $\tilde{\rho}<1$. When $N$ is small to
the extent that inequality $\tilde{\rho}<1$
becomes the stricter condition, mean delay (resp. delay jitter) is guaranteed to be finite as long as the system is not saturated.
In this case, $\tau_{BBMD}$ (resp. $\tau_{BBDJ}$)$=\tau_s$ and
$S_{BBMD}$ (resp. $S_{BBDJ}$)$=S_s$. Since the behavior of MPR WLANs
at $(\tau_s, S_s)$ has been extensively studied in another paper of
ours \cite{Zhang}, we focus our interest in the large $N$ case in
this paper.

\subsection{SBMD and SBDJ throughput}
Scenarios 3 and 4 in Fig. \ref{fig:4_scenarios} imply that it is
theoretically possible to operate the system at a throughput higher
than $S_s$ while maintaining $p_c<1/r^2$ (or $p_c<1/r^3$ for BDJ).
This is the case only if the long-term average output rate can be
maintained at the higher throughput. In practice, however, it is not
safe to load the system with an input rate higher than $S_s$. To see
this, we note that because of the random packet arrival, it is always possible for the
system to evolve to a state where all queues are backlogged at certain point of time. Once this persists for a while, the system will behave as if it is saturated and the throughput will degenerate to the saturation throughput. If the offered load $N\lambda$ is set to be higher than the saturation throughput, the backlog will continue to build up and the system will never get out of saturation again. The delay will then go to infinity. The intricacy lies in the fact that we have a system in which the service rate can be degraded ($p_c$
increases) once saturation sets it, which is unlike an ``ordinary" queueing system in which the service rate is independent of the system state. This particular aspect of $p_c$ transiting to a higher value under saturation is not captured in our $M/G/1/V_m$ queue model, which looks at a tagged queue and assumes a constant $p_c$.

\begin{figure}
\centering
\includegraphics[width=0.5\textwidth]{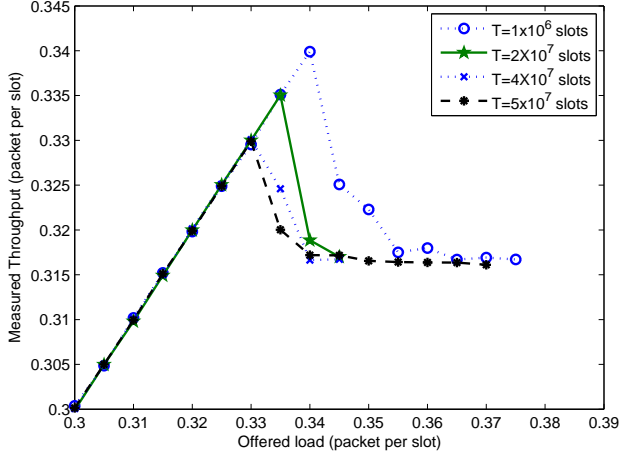}
\caption{Throughput vs. Offered load for different simulation lengths.}
 \label{fig:unstable}
\end{figure}

This phenomenon is illustrated in Fig. \ref{fig:unstable}, where we have simulated an ALOHA-like system with $N=50$ and slot lengths $T_{succ}=T_{coll}=T_{idle}=H+PL/\text{data rate}$. Here $H$ is the transmission time of PHY header and MAC header. In the figure, throughput is plotted as a function of offered load. Each point in the figure is a result of a simulation run of $T$ slots. Different curves in the figure correspond to simulation runs with different numbers of simulated time slots. As shown, when the offered load exceeds the saturation throughput (around 0.316 packets/slot), it may be sustainable for a short while (in short simulation runs), but eventually becomes unsustainable as time progresses (in long simulation runs).

Define safe BMD throughput $S_{SBMD}$ and safe BDJ throughput
$S_{SBDJ}$ to be the highest throughput that can be \emph{safely}
sustained with bounded mean delay and delay jitter, respectively.
Based on the above articulation,
\begin{equation}\label{eq:S_SBMD}
    S_{SBMD}=\min(S_{BBMD}, S_s)
\end{equation}
and
\begin{equation}\label{eq:S_SBDJ}
   S_{BBDJ}=\min(S_{SBDJ}, S_s)
\end{equation}

\subsection{Super-linear scaling of maximum SBMD and SBDJ throughput}
Given $M$, $S_{SBMD}$ and $S_{SBDJ}$ can be maximized by deploying an optimal $r$ in the EB process. Denote the maximum $S_{SBMD}$ and $S_{SBDJ}$ by $S^*_{SBMD}(M)$ and $S^*_{SBDJ}(M)$, respectively. The corresponding optimal $r's$ are denoted by $r^*_{SBMD}(M)$ and $r^*_{SBDJ}(M)$, respectively.

In our earlier work \cite{Zheng:06ICC, Zheng:06WLN, Zhang}, we have proved that the maximum saturation throughput of MPR WLANs increases \emph{super-linearly} with MPR capability $M$. That is, $S_s^*/M$ increases with $M$.  In this subsection, we will show that super-linear scaling also holds for the maximum SBMD and SBDJ throughput. That is, $\frac{S^*_{SBMD}(M)}{M}$ and $\frac{S^*_{SBDJ}(M)}{M}$ increases with $M$.  In practice, MPR capability comes with a cost. For example, $M$ could be the number of antennas at AP for multi-antenna WLAN or the spectrum spreading factor in CDMA WLAN. Super-linear throughput scaling implies that the throughput \emph{per unit cost} grows with $M$. This provides an strong incentive to implement MPR in WLAN.
\medskip

\noindent\emph{The large N case}

To find the optimal $r$, we need to solve the following problems.
\begin{eqnarray}
    r^*_{SBMD}(M)&=&\arg\max_r S_{SBMD}(M)\\
    &=&\arg\max_r \min (S_{BBMD}(M), S_s(M))\nonumber
\end{eqnarray}
and
\begin{eqnarray}
    r^*_{SBDJ}(M)&=&\arg\max_r S_{SBDJ}(M)\\
    &=&\arg\max_r \min (S_{BBDJ}(M), S_s(M))\nonumber
\end{eqnarray}
respectively.
As discussed in Section V-A, $S_{BBMD}$ and $S_{BBDJ}$ differ from $S_s$ in the large $N$ case.  Therefore,  the optimal solutions to the above problems occur when
\begin{equation}\label{eq:S_BBMDeqS_s}
S_{BBMD}(M)=S_s(M)
\end{equation}
and
\begin{equation}\label{eq:S_BBDJeqS_s}
S_{BBDJ}(M)=S_s(M),
\end{equation}
respectively. In other words, the optimal solutions correspond to somewhere between scenario 2 and scenario 3 shown in Fig. \ref{fig:4_scenarios}.

To solve \eqref{eq:S_BBMDeqS_s}, rewrite the equation into \eqref{eq:45}, which can be further rewritten into \eqref{eq:45a}. (\eqref{eq:45} and \eqref{eq:45a} are shown at top of next page.)

\begin{table*}[t]
\begin{eqnarray}\label{eq:45}
\frac{PL\sum_{k=1}^Mk\binom{N}{k}\tau_{BBMD}^k(1-\tau_{BBMD})^{N-k}}{P_{idle}^G(\tau_{BBMD})T_{idle}+P_{coll}^G(\tau_{BBMD})T_{coll}+P_{succ}^G(\tau_{BBMD})T_{succ}}
=\frac{PL\sum_{k=1}^Mk\binom{N}{k}\tau_{s}^k(1-\tau_s)^{N-k}}{P_{idle}^G(\tau_{s})T_{idle}+P_{coll}^G(\tau_{s})T_{coll}+P_{succ}^G(\tau_{s})T_{succ}},
\end{eqnarray}

\begin{eqnarray}\label{eq:45a}
\frac{N\lambda\sum_{k=0}^{M-1}\binom{N-1}{k}\tau_{BBMD}^k(1-\tau_{BBMD})^{N-1-k}}{P_{idle}^G(\tau_{BBMD})T_{idle}+P_{coll}^G(\tau_{BBMD})T_{coll}+P_{succ}^G(\tau_{BBMD})T_{succ}}
=\frac{N\lambda\sum_{k=0}^{M-1}\binom{N-1}{k}\tau_{s}^k(1-\tau_s)^{N-1-k}}{P_{idle}^G(\tau_{s})T_{idle}+P_{coll}^G(\tau_{s})T_{coll}+P_{succ}^G(\tau_{s})T_{succ}}.
\end{eqnarray}
\end{table*}
For large $N$, $p_c\rightarrow1/r$ when $\tau\rightarrow\tau_s$ and
$p_c\rightarrow1/r^2$ when $\tau\rightarrow\tau_{BBMD}$. That is,
\begin{equation}\label{eq:46}
    \sum_{k=0}^{M-1}\binom{N-1}{k}\tau_{s}^k(1-\tau_s)^{N-1-k} =1-1/r,
\end{equation}
and
\begin{equation}\label{eq:47}
    \sum_{k=0}^{M-1}\binom{N-1}{k}\tau_{BBMD}^k(1-\tau_{BBMD})^{N-1-k} =1-1/r^2,
\end{equation}
Substituting \eqref{eq:46}-\eqref{eq:47} to \eqref{eq:45a}, $r^*_{SBMD}(M)$ can be solved numerically. Likewise, we can solve for $r^*_{SBDJ}$  by replacing $1/r^2$ with $1/r^3$ in the above equations.
\\
\\
\noindent\emph{The small $N$ case}

When $N$ is small to the extent that $\tilde{\rho}<1$ is a stricter condition than $p_cr^2<1$ (or $p_cr^3<1$), $(\tau_{BBMD},S_{BBMD})$ (or
$(\tau_{BBDJ},S_{BBDJ})$) overlaps with $(\tau_s,S_s)$. Hence,
$S^*_{SBMD}(M)$ (or $S^*_{SBDJ}(M)$) is equal to ${S_s^*(M)}$. In our earlier work in
\cite{Zhang} we have proved that ${S_s^*(M)}$ scales super-linearly
with $M$. Hence, super-linear scaling of $S^*_{SBMD}(M)$ or
$S^*_{SBDJ}(M)$ is straightforward in this case.
\medskip

\begin{figure}[t]
\centering
\includegraphics[width=0.5\textwidth]{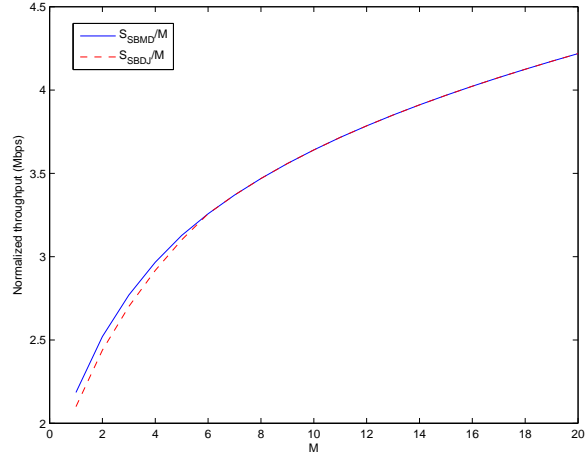}
\caption{Super-linear throughput scaling when $N=50$ and $T_{succ}=T_{coll}=T_{idle}$.}
 \label{fig:superlinear_ALOHA}
\end{figure}

In Fig. \ref{fig:superlinear_ALOHA}, $\frac{S^*_{SBMD}(M)}{M}$ and $\frac{S^*_{SBDJ}(M)}{M}$ are plotted against $M$ when $N=50$ and $T_{succ}=T_{coll}=T_{idle}=H+PL/\text{data rate}$, where $H$ is the transmission time of PHY header and MAC header. Detailed parameters and values are listed in Table 1. This setting corresponds to ALOHA-like systems where slot length does not vary with channel status.  The figure shows that SBMD and SBDJ throughput scale super-linearly with $M$: normalized throughput $\frac{S^*_{SBMD}(M)}{M}$ and $\frac{S^*_{SBDJ}(M)}{M}$ increase with $M$. This result, together with our earlier work \cite{Zhang}, provides a strong incentive to deploy MPR in future WLANs, no matter whether the underlying application is delay sensitive or not.

\begin{table}[!t]\label{Table1}
\caption{System Parameters} \label{system_param} \centering
\begin{tabular}{|l|l|}
\hline
\textbf{Parameter} & \textbf{Value} \\
\hhline{|=|=|} PHY Header  &20 $\mu$s \\
MAC Header  &244 bits transmitted at 6 Mbps \\
Data Transmission Rate &6 Mbps \\
CWmin   &16 \\
CWmax   &Inf \\
Retry Limit &Inf\\
DIFS    &34 $\mu$s \\
SIFS &16 $\mu$s \\
Mini slot length &9 $\mu$s \\
ACK &112 bits \\
Packet Size &1000 bytes (8184 bits)\\
\hline
\end{tabular}
\end{table}

\begin{figure}[t]
\centering
\includegraphics[width=0.5\textwidth]{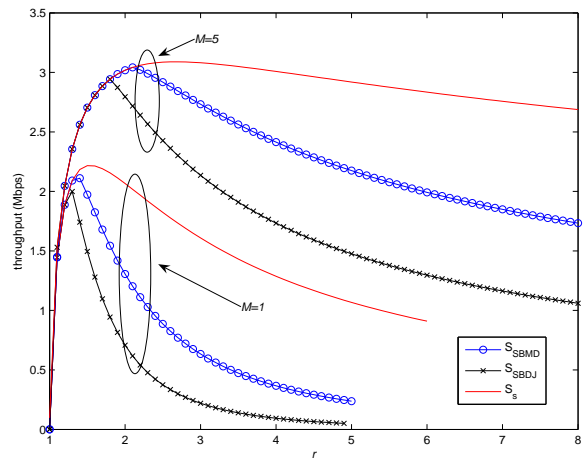}
\caption{Throughput vs. $r$ when $N=50$ and $T_{succ}=T_{coll}=T_{idle}$.}
 \label{fig:S_vs_r_ALOHA}
\end{figure}

It is worth noting that SBMD and SBDJ throughputs are more sensitive to $r$ than saturation throughput is, as shown in Fig. \ref{fig:S_vs_r_ALOHA}. It can be seen that depending on $M$, the commonly adopted binary EB, where $r=2$, can be far from optimum. Therefore, one should be more careful in choosing the right $r$ to avoid severe degradation in sustainable throughput when delay is a concern.

A close observation of Fig. \ref{fig:S_vs_r_ALOHA} shows that large MPR capability $M$ decreases the sensitive of throughput to $r$. This provides another incentive to deploy MPR in WLANs, because the system is now more robust against mis-selection of $r$.

\begin{figure}[t]
\centering
\includegraphics[width=0.5\textwidth]{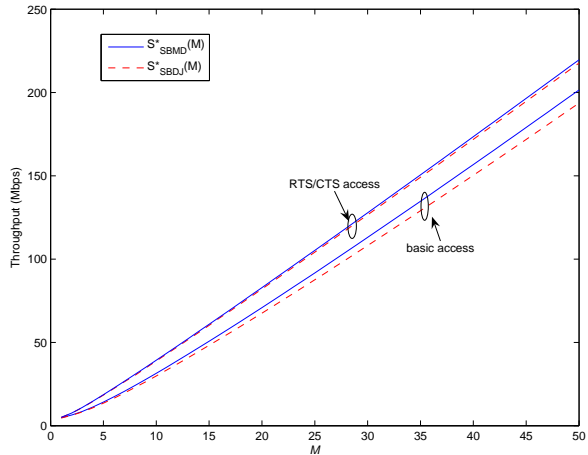}
\caption{Throughput vs. $M$ in basic-access and RTS/CTS-access modes.}
 \label{fig:S_vs_M_CSMA}
\end{figure}

\begin{figure}[t]
\centering
\includegraphics[width=0.5\textwidth]{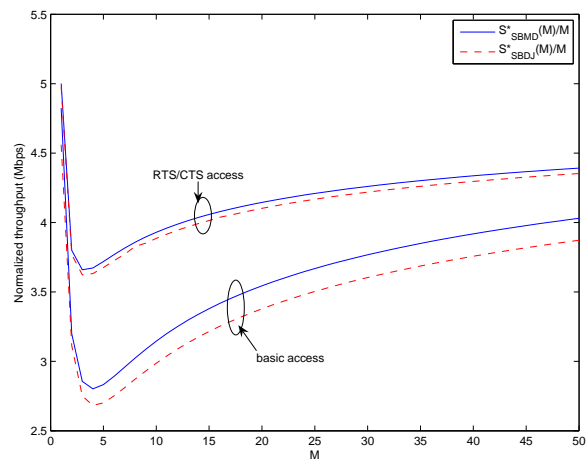}
\caption{Super-linear throughput scaling in basic-access and RTS/CTS-access modes.}
 \label{fig:superlinear_CSMA}
\end{figure}

Super-linear throughput scaling is also observed when slot lengths are set according to DCF standard. In Fig. \ref{fig:S_vs_M_CSMA} and Fig.
\ref{fig:superlinear_CSMA}, throughput and normalized throughput are
plotted for DCF basic-access and RTS/CTS access modes, where slot lengths are given by the following equations, respectively.
\begin{equation}\label{eq:T-basic}
    \begin{cases}
    T_{idle}^{\text{basic mode}}=\sigma\\
    T_{coll}^{\text{basic mode}}=H+PL/\text{data rate}+DIFS\\
    T_{succ}^{\text{basic mode}}=H+PL/\text{data rate}+SIFS+ACK+DIFS
    \end{cases}
\end{equation}
and
\begin{equation}\label{eq:T-rts}
    \left\{\begin{aligned}
    T_{idle}^{\text{RTS/CTS}}=&\sigma\\
    T_{coll}^{\text{RTS/CTS}}=&RTS+DIFS\\
    T_{succ}^{\text{RTS/CTS}}=&RTS+CTS+H+PL/\text{data rate}\\&+3SIFS+ACK+DIFS
    \end{aligned}\right.
\end{equation}
where $\sigma$ is the time needed for a station to detect the packet transmission from any other station and is typically much smaller than $T_{coll}$ and $T_{succ}$; \textit{ACK} is the transmission time of an ACK packet; $\delta$ is the propagation delay; and \textit{SIFS} and \textit{DIFS} are the inter-frame space durations.

The figures show that SBMD and SBDJ throughputs are greatly improved
due to the MPR enhancement in the PHY layer. Moreover, $\frac{S^*_{SBMD}(M)}{M}$ and $\frac{S^*_{SBDJ}(M)}{M}$ increase with $M$ for both access modes when $M$ is relatively large.

\section{Numerical Results}
In this section, we further illustrate the results in Section IV and Section V through two examples: systems corresponding to scenario 1 and 4 as defined in Fig. \ref{fig:4_scenarios}.

\subsection*{Scenario 1}
In this subsection, we set $T_{succ}=T_{coll}=T_{idle}=H+PL/\text{data rate}$, $r=2$, $M=1$ and $N=50$. A simple calculation shows that $\tau_{BBDJ}<\tau_{BBMD}<\tau_s\leq\tau^*$ and $S_{BBDJ}<S_{BBMD}\leq S_s$, corresponding to scenario 1 in Fig. \ref{fig:4_scenarios}. In this scenario, it is only possible to operate the system at an
offered load lower than $S_{BBMD}$, and there is only one possible operating point, $\tau_l$, for each offered load $N\lambda$.

\begin{figure}[h]
\centering
\includegraphics[width=0.5\textwidth]{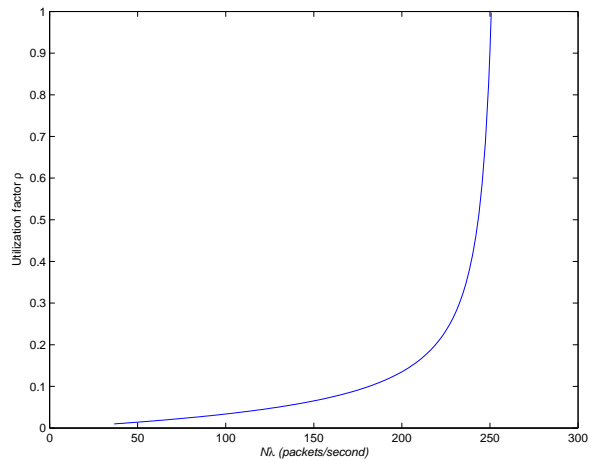}
\caption{Utilization factor vs. offered load.}
 \label{fig:utility_ALOHA}
\end{figure}

In Fig. \ref{fig:utility_ALOHA}, the utilization factor $\rho$ is plotted against offer load $N\lambda$. Unlike traditional queueing systems where $\rho$ increases at the same rate as offered load, the figure shows that $\rho$ increases much faster than $N\lambda$ in WLANs, especially when $N\lambda$ is large. This is due to the fact
that as $\lambda$ increases, not only does the input rate of the tagged queue increases, the mean service time $\mathrm{E}[X_{ne}]$ also increases due to heavier contention among nodes.  Beyond certain point, the system approaches saturation (i.e., $\rho=1$) very rapidly.

\begin{figure}[h]
\centering
\includegraphics[width=0.5\textwidth]{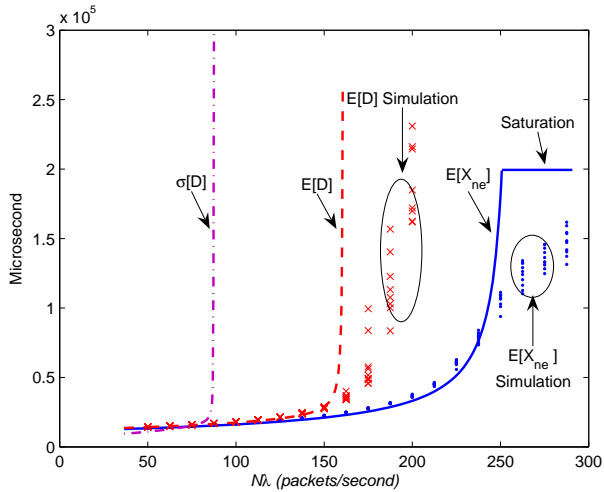}
\caption{Delay vs. offered load.}
 \label{fig:delay_vs_load_ALOHA}
\end{figure}
In Fig. \ref{fig:delay_vs_load_ALOHA}, we plot $\mathrm{E}[X_{ne}]$, $\mathrm{E}[D]$, and $\sigma[D]=\sqrt{\mathrm{VAR}[D]}$ against offered load $N\lambda$. The solid lines represent the results obtained from analysis. The markers correspond to simulation results. It is not surprising that when the offered load reaches the saturation throughput (which is also the point at which $\rho$ goes to 1 in Fig. \ref{fig:utility_ALOHA}), $\mathrm{E}[X_{ne}]$ quickly converges to a constant equal to the reciprocal of the saturation throughput of one
user. As predicted by the analysis, mean packet delay $\mathrm{E}[D]$ becomes infinite earlier than $\mathrm{E}[X_{ne}]$ because $S_{BBMD}<S_s$. Likewise, the offered load that can be sustained with finite delay jitter is even lower: $\sigma[D]$ approaches infinity earlier than $\mathrm{E}[D]$. In this scenario, it is necessary to load the system far below the saturation throughput to guarantee finite delay and delay jitter.

In this figure, we have conducted several independent simulation experiments to measure packet delay. The simulations are conducted with MATLAB7.1. One interesting observation is that different simulation experiments do not yield the same results when offered load is relatively high, even if we run each experiment for a long time (at the order of hours). This is, however, not surprising, considering the following facts. When offered load is higher than $S_{BBMD}$, $\mathrm{E}[X_{ne}^2]$ is infinite, and so is $\mathrm{VAR}[X_{ne}]$. Hence, sample mean of $X_{ne}$ obtained from numerical simulation will not converge to the true mean $\mathrm{E}[X_{ne}]$
no matter how much data are collected. Likewise, when $\sigma[D]$ is infinite, the simulation results for mean packet delay do not converge. This phenomenon, referred to as \emph{immeasurability}, has not been discovered in previous work, although a close look at the simulation results in some previous work (e.g., \cite{Garetto:05}) does reveal inconsistency of independent simulation runs under the same system parameters. Interested readers are referred to \cite{Soung}, where we discuss the immeasurability issue in more depth.

\subsection{Scenario 4}
In this subsection, we set slot lengths according to \eqref{eq:T-basic} and let $r=2$, $M=1$ and $N=50$. Simple calculation shows that $\tau^*<\tau_{BBDJ}<\tau_{BBMD}<\tau_s$ and $S_s<S_{BBDJ}<S_{BBMD}$ in this case, which corresponds to scenario 4 in Fig. \ref{fig:4_scenarios}.

\begin{figure}
\centering
\includegraphics[width=0.5\textwidth]{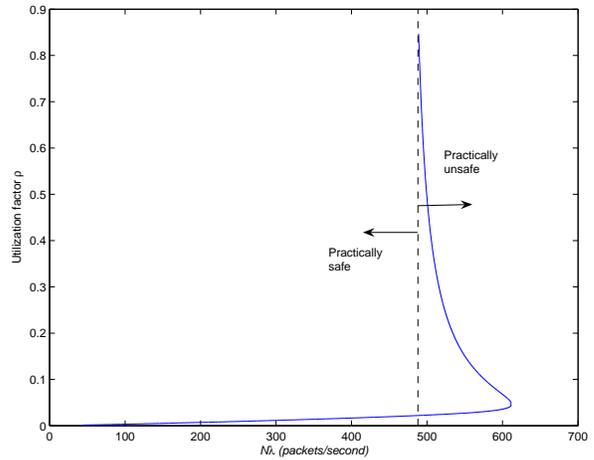}
\caption{Utilization factor vs. offered load in WLANs with
basic-access mode.}
 \label{fig:utility_basic}
\end{figure}

It is easy to see from Fig. \ref{fig:4_scenarios} that in scenario 4, $\tau_l$ is the only operating point when input traffic $N\lambda<S_s$. When $N\lambda>S_s$, however, both $\tau_l$ and $\tau_r$ are possible operating points in theory. In other words, an offered load $S_s<N\lambda<S^*$ can result in two attempt rates under non-saturation condition. $\tau_l$ corresponds to a lower contention level, while $\tau_r$ leads to a higher contention level.
This is illustrated in Fig. \ref{fig:utility_basic}, where $\rho$ is plotted against $N\lambda$. It can be seen that when $N\lambda$ is larger than $S_s$, which is around 486.5 packets per second, there are two $\rho$'s corresponding one $N\lambda$. The smaller $\rho$ results from $\tau_l$ and the larger one results from $\tau_r$. If the system operates at $\tau_r$, it reaches saturation when $N\lambda$ approaches $S_s$.

\begin{figure}
\centering
\includegraphics[width=0.5\textwidth]{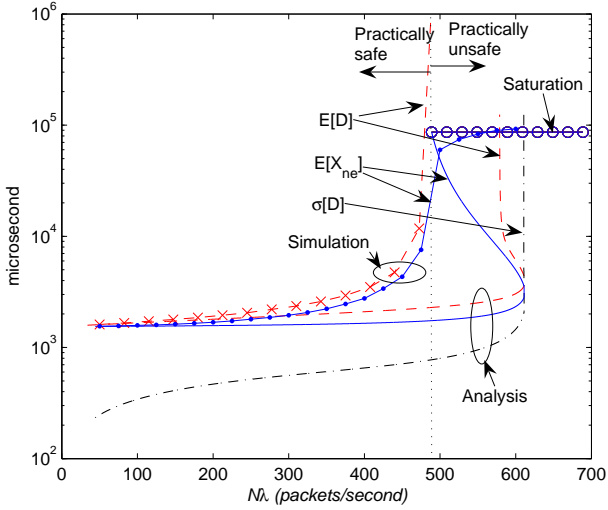}
\caption{Delay vs. offered load in WLANs with basic-access mode.}
 \label{fig:delay_vs_load_basic}
\end{figure}

In Fig. \ref{fig:delay_vs_load_basic}, $\mathrm{E}[X_{ne}]$, $\mathrm{E}[D]$, and $\sigma[D]=\sqrt{\mathrm{VAR}[D]}$ are plotted against $N\lambda$. The curves without marks represent the results obtained from analysis. Similar to the arguments in Fig. \ref{fig:utility_basic}, there are two $\mathrm{E}[X_{ne}]$'s corresponding to one $N\lambda$ when $N\lambda>S_s$. When saturated, $\mathrm{E}[X_{ne}]$ is equal to the reciprocal of the saturation throughput of one station. Likewise, ``kinks" in $\mathrm{E}[D]$ and $\sigma[D]$ are also observed when $N\lambda>S_{BBMD}$ and
$N\lambda>S_{BBDJ}$, respectively.  (Note that the ``kink" in $\sigma[D] $ is not obvious because $S_{BBDJ}$ is close to $S^*$ in this example).

As discussed in Section V-B, it is not safe to load the system with an offered load higher than $S_s$ in practice. Otherwise, system throughput will eventually collapse to $S_s$ and packet delay will go to infinity. To see this, we plot packet lengths measured by simulations in Fig. \ref{fig:delay_vs_load_basic}.  As expected, we are unable observe a throughput higher than $S_s$ in the simulations. When offered load $N\lambda$ approaches $S_s$, the mean service time quickly converges to the reciprocal of saturation throughput, implying that the system is already saturated. In the meantime, packet delay becomes unbounded as well.

Note that unlike scenario 1, numerical results from different simulation runs converge in Fig. \ref{fig:delay_vs_load_basic}. This is because for the region of bounded mean delay, the variance of delay does not go to infinity.

\section{Discussions}
Up to this point, we have assumed that there are no restrictions on the maximum retry limit and the maximum contention window. In this section, we show that the conclusions reached in the preceding sections still hold even when there is a retry limit $K$ or a maximum contention window $CW_{max}$. That is, it is necessary to have $p_cr^n<1$ if delay moments are to be kept small. For example, we need $p_c$ to be smaller than $1\big/r^2$ if a small mean packet delay $E[D]$ is desired.

\subsection{The case with a retry limit $K$}
The system becomes a lossy one in the presence of a retry limit $K$. Packets that cannot go through after $K$ retrials (i.e., $K+1$ transmissions including the initial attempt) are dropped from the system. Hence, $K$ should be reasonably large to keep the packet loss rate low. The typical value of $K$ is about 4 to 7 for IEEE 802.11 standards.

\begin{figure}
\centering
\includegraphics[width=0.5\textwidth]{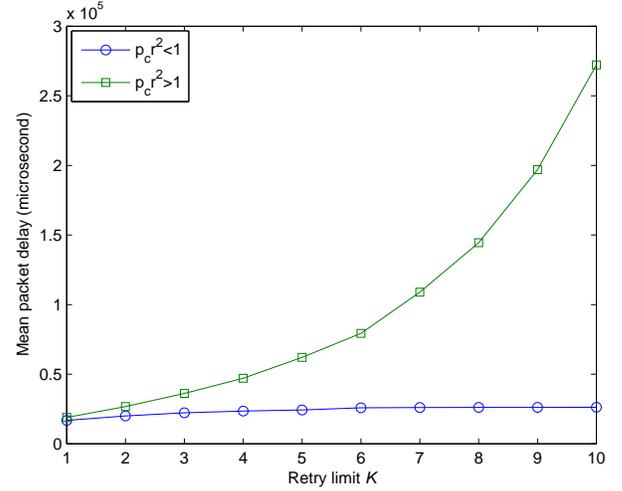}
\caption{Mean packet delay vs. retry limit.}
 \label{fig:K}
\end{figure}

With retry limit $K$, Eqns. \eqref{eq:succ-j}, \eqref{Xne-moment}, and \eqref{term} are modified as follows. The probability that a packet that eventually gets through is successfully transmitted on its $j^{th}$ transmission is given by
\begin{equation}
\Pr\{R=j\}=\frac{p_c^{j-1}(1-p_c)}{1-p_c^{K+1}}~\forall 1\leq j\leq K+1.
\end{equation}
Hence,
\begin{eqnarray}
&&\mathrm{E}[X_{ne}^n]\nonumber\\
&=&\frac{1-p_c}{1-p_c^{K+1}}\sum^{K+1}_{j=1}p_c^{j-1}\mathrm{E}\big[\big(\sum^j_{i=1}C_i+(j-1)T_{coll}+T_{succ}\big)^n\big]\nonumber\\
&=&\frac{1-p_c}{1-p_c^{K+1}}\sum^{K+1}_{j=1}p_c^{j-1}\bigg(\mathrm{E}[C_j^n]+\text{other terms with}\nonumber\\
&&\text{power of $r$ lower than}~(j-1)n\bigg).
\end{eqnarray}
It is easy to see that $\mathrm{E}[X_{ne}^n]$ contains the term
\begin{eqnarray}\label{K-term}
A_1^n\frac{W_0^n(1-p_c)}{(n+1)(1-p_c^{K+1})}\sum^{K+1}_{j=1}p_c^{j-1}r^{(j-1)n}.
\end{eqnarray}
With retry limit $K$, $\mathrm{E}[X_{ne}^n]$ is always bounded because \eqref{K-term} is a summation of a finite number of terms. However, a closer look at \eqref{K-term} indicates that it is a summation of increasing terms when $p_cr^n>1$. That is, $p_c^{j-1}r^{(j-1)n}$ grows with $j$. In this case, packet delay can be quite large even with a moderate $K$. On the other hand, \eqref{K-term} is a summation of diminishing terms when $p_cr^n<1$, and hence does not grow noticeably with $K$. This is illustrated in Fig. \ref{fig:K} where mean packet delay is plotted as a function of retry limit $K$ when the simulation settings are the same as those in scenario 1 in Section VI. It can be seen that when $p_cr^2<1$, mean packet delay is maintained at a low level regardless of $K$. In contrast, mean packet delay grows rapidly with $K$ when $p_cr^2>1$. When $K=7$, for example, packet delay in the case of $p_cr^2>1$ is more than four times that of the case when $p_cr^2<1$.

\subsection{The case with $CW_{max}$}

In the presence of $CW_{max}$, the saturation throughput degrades quickly due to the excessive collision when the number of users $N$ exceeds $1/CW_{max}$. In other words, the system is easy to be saturated by a low input traffic rate when $N$ is large, and all moments of packet delay will go to infinity in this case.

When the network is unsaturated due to small $N$ or low input traffic rate, the derivations in Section IV.A can be slightly modified as follows to take $CW_{max}$ into consideration.
\begin{equation}
\Pr\{B_i=k\}=
\begin{cases}
\frac{1}{r^{i-1}W_0} & \forall k\in[0,r^{i-1}W_0-1]\\
&\text{if}~i\leq 1+\log_r\frac{CW_{max}}{W_0}\\
\frac{1}{CW_{max}} & \forall k\in[0,CW_{max}-1]\\
&\text{if}~i> 1+\log_r\frac{CW_{max}}{W_0}
\end{cases}
\end{equation}
\begin{eqnarray}
\mathrm{E}[C_i^n]&=&(-1)^n\frac{d^n\hat{B}_i(L^*(s))}{ds^n}\bigg|_{s=0}\nonumber\\
&=&
\begin{cases}
A_1^n\frac{r^{(i-1)n}W_0^n}{n+1} & \text{if}~i\leq 1+\log_r\frac{CW_{max}}{W_0}\\
A_1^n\frac{CW_{max}}{n+1} & \text{if}~i> 1+\log_r\frac{CW_{max}}{W_0}
\end{cases}
\end{eqnarray}
Consequently, $\mathrm{E}[X_{ne}^n]$ contains the item
\begin{eqnarray}\label{CWmax-term}
&&\frac{A_1^n}{n+1}\bigg((1-p_c)W_0^n\sum_{j=1}^{1+\log_r\frac{CW_{max}}{W_0}}p_c^jr^{(i-1)n}\nonumber\\
&&+CW_{max}^n\sum_{2+\log_r\frac{CW_{max}}{W_0}}^\infty p_c^j\bigg)\nonumber\\
&=&\frac{A_1^n}{n+1}\bigg((1-p_c)W_0^n\sum_{j=1}^{1+\log_r\frac{CW_{max}}{W_0}}p_c^jr^{(i-1)n}\nonumber\\
&&+CW_{max}^np_c^{2+\log_r\frac{CW_{max}}{W_0}}\bigg).
\end{eqnarray}
Similar to the case with retry limit $K$, \eqref{CWmax-term} implies that $\mathrm{E}[X_{ne}^n]$ is always finite whenever the network is unsaturated. However, $p_cr^n$ is still a determining factor for packet delay. When $p_cr^n>1$, $p_c^jr^{(j-1)n}$ increases with $j$, implying that \eqref{CWmax-term} contains a summation of increasing terms and hence grows rapidly. In other words, the delay moments, although finite, may be unfavorably large. Hence, it is still desired to keep $p_cr^n<1$ even if there is a maximum contention window $CW_{max}$.

\section{Conclusions}
In this paper, we have investigated how much throughput can be sustained subject to bounded mean delay and delay jitter requirements in MPR WLANs under non-saturation condition. Using an $M/G/1/V_m$ queueing model, we have derived an explicit
expression for the distribution (in transform) of packet delay. The analysis establishes sufficient and necessary conditions for mean delay and delay jitter to be bounded: $\tilde{\rho}<1$ and $p_c<1/r^2$ for
bounded mean delay; and $\tilde{\rho}<1$ and $p_c<1/r^3$ for bounded delay
jitter, respectively. This result implies that the mean packet delay and delay jitter can go to infinity even if the system is not saturated. Based on the analysis, we define SBMD and SBDJ throughputs to be the maximum throughput that can safely guarantee bounded mean delay and delay jitter. These are arguably more sensible definitions of throughput for delay-sensitive applications. To maximize SBMD (resp. SBDJ) throughputs, the backoff factor $r$
should be carefully chosen for given $M$ and slot lengths. Our results show that under some circumstances, the widely adopted binary EB where $r=2$ yields a throughput that is far from optimum.

Together with our previous work on MPR WLANs, this paper has completed the demonstration of MPR as a powerful
capacity-enhancement technique for both delay-sensitive and delay-tolerant applications. Firstly, the maximum SBMD and SBDJ throughputs are shown to scale super-linearly with MPR capability $M$. That is, throughput per unit cost increases with $M$ in MPR WLANs. Secondly, the sensitivity of SBMD and SBDJ throughputs with respect to backoff factor $r$ decreases for large $M$. This implies that an MPR system is more robust against suboptimality in the selection of $r$.

In the paper, we have demonstrated the ``unsafeness" of loading the system with a traffic load higher than saturation throughput, even when the sufficient and necessary condition for bounded delay moments is satisfied. An interesting future research direction is to devise mechanisms to ``safely" sustain a throughput higher than saturation throughput. The simulation results in \cite{Duffy:05} suggested that limiting buffer size to a small value could be an effective approach to convert an unsafe throughput to a safe one. Alternatively, regulating  input traffic is another possible approach. The key idea in the design of ``safening" mechanisms is to allow buffers to be empty from time to time so that the system remains unsaturated.

\appendices
\section{Proof of Theorem 1}
To prove the ``if" part of Theorem \ref{theorem1}, we show that $\rho<1$ when $\tau<\tau_s$. To prove the ``only if" part of Theorem \ref{theorem1}, we show that when $\tau>\tau_s$, $\rho$ exceeds 1, which
violates the fact that $\rho$ must be smaller than or equal to 1.

Before proving Theorem \ref{theorem1}, we present the following lemma. To avoid
lengthy derivation, we focus on EB-based WLANs with
$M=1$ and $T=T_{succ}=T_{coll}=T{idle}$ in Lemma \ref{lemma:rho}. The lemma, however, can be generalized to other
scenarios.
\begin{lemma}\label{lemma:rho}
$\tilde{\rho}=\lambda \mathrm{E}[X_{ne}]$ is an increasing function of $\tau$  in
EB-based WLANs where $T=T_{coll}=T_{succ}=T_{idle}$.
\end{lemma}
\begin{IEEEproof}
In this case, $\mathrm{E}[X_{ne}]=T\frac{W_0(1-p_c)+(1-rp_c)}{2(1-rp_c)(1-p_c)}$.
It is obvious that $\mathrm{E}[X_{ne}]$ increases with $p_c$, and hence is an
increasing function of  $\tau$.
\\
Case (i) $\tau<\tau^*$: In this case, it is trivially
straightforward that $\tilde{\rho}=\lambda\mathrm{E}[X_{ne}]$ is an increasing
function of $\tau$, as $\lambda$ also increases with $\tau$.
\\
Case (ii) $\tau\geq\tau^*$: In this case, $\lambda$ decreases with
$\tau$. That is, $\frac{\mathrm{d}\lambda}{\mathrm{d}\tau}<0$, which leads to $N\lambda>1$ by considering \eqref{eq:S-unsat}.

The derivative of $\tilde{\rho}$ with respect to $\tau$ is calculated as
\begin{eqnarray}\label{eq:app-4}
      &&\frac{\partial\tilde{\rho}}{\partial\tau}= \frac{\partial\left( \lambda \mathrm{E}[X_{ne}] \right)}{\partial\tau}\nonumber\\
        &=&PL\frac{\partial}{\partial\tau}\left(\frac{\tau(W_0+r)(1-\tau)^{N-1}+\tau(1-r)}{2(1-r+r(1-\tau)^{N-1})} \right)\nonumber\\
      &=&PL\bigg(\frac{W_0r(1-\tau)^{2N-2}+W_0(1-\tau)^{N-2}(1-N\tau)(1-r)}
      {2\left( r(1-\tau)^{N-1}+1-r \right)^2}\nonumber\\
      &&+\frac{2r(1-\tau)^{N-1}(1-p_cr)+(1-r)^2}{2\left( r(1-\tau)^{N-1}+1-r \right)^2}\bigg)
\end{eqnarray}
Since $p_c<1/r$ (which is necessary for steady state), $N\tau>1$,
and $r\geq1$, it is easily seen that
$\frac{\partial\tilde{\rho}}{\partial\tau}>0$. Hence $\tilde{\rho}$ is an increasing
function of $\tau$.
\end{IEEEproof}
When $\tau=\tau_s$, $\tilde{\rho}=1$.  Lemma \ref{lemma:rho} implies that $\tilde{\rho}> 1$
if $\tau>\tau_s$ and $\tilde{\rho}< 1$ if $\tau<\tau_s$. As discussed in Section IV-C, $\tilde{\rho}<1$ implies $\rho<1$ and the network is unsaturated. Likewise,the system cannot be stable when $\tilde{\rho}$ exceeds 1. This completes the proof.
$\hfill\blacksquare$

\section{Expressions for $\theta_1$, $\theta_2$, and $\theta_3$ in \eqref{eq:EXne3Conv}}
\begin{eqnarray}\label{eq:theta_1}
      \theta_1 &=& \frac{A_1^3}{4}\left( -\frac{W_0^2}{1-r^2p_c} +\frac{4W_0}{1-rp_c}- \frac{3}{1-p_c} \right)\nonumber\\
      &+& \frac{A_1A_2}{4}\left( \frac{W_0^2}{1-r^2p_c} - \frac{6W_0}{1-rp_c} + \frac{5}{1-p_c} \right) \nonumber\\
        &+&\frac{A_3}{2}\left( \frac{W_0}{1-rp_c}- \frac{1}{1-p_c} \right)
\end{eqnarray}
\begin{equation}\label{eq:theta_2}
    \begin{split}
      &\theta_2= \frac{A_1^3}{12}\begin{pmatrix}
        \frac{W_0^3(1-r^3p_c^2)}{2(1-rp_c)(1-r^2p_c)(1-r^3p_c)}
        - \frac{3W_0^2(1+rp_c)}{(1-rp_c)(1-r^2p_c)} \\
        + \frac{11W_0(1-rp_c^2)}{2(1-p_c)(1-rp_c)^2}
        -\frac{W_0^2(1-r^2p_c^2)}{2(1-p_c)(1-r^2p_c)^2} - \frac{5(1+p_c)}{2(1-p_c)^2} \end{pmatrix} \\
      &+\frac{A_1^2}{12}\begin{pmatrix} T_{coll} \bigg( \frac{W_0^2p_c(1+r^2-2r^2p_c)}{(1-p_c)(1-r^2p_c)^2}
        -\frac{6W_0p_c(1+r-2rp_c)}{(1-p_c)(1-rp_c)^2} \\+\frac{10p_c}{(1-p_c)^2}
        \bigg)
        +T_{succ} \bigg( \frac{W_0^2}{1-r^2p_c}-\frac{6W_0}{1-rp_c}
        +\frac{5}{1-p_c}\bigg)\end{pmatrix} \\
        &+A_1A_2\begin{pmatrix}\frac{W_0^2}{4}\frac{1+rp_c}{(1-rp_c)(1-r^2p_c)}
        -\frac{W_0}{2} \frac{1-rp_c^2}{(1-p_c)(1-rp_c)^2 }\\+
        \frac{1+p_c}{4(1-p_c)^2}\end{pmatrix}\\
        &+A_2\begin{pmatrix} \frac{T_{coll}p_c}{2}\frac{W_0(1+r-2rp_c)(1-p_c)-2(1-rp_c)^2}{(1-p_c)^2(1-rp_c)^2}\\
        + \frac{T_{succ}}{2}\frac{W_0(1-p_c)-(1-rp_c)}{(1-p_c)(1-rp_c)}\end{pmatrix}
    \end{split}
\end{equation}
\begin{equation}\label{eq:theta_3}
    \begin{split}
        & \theta_3=A_1^3\begin{pmatrix}\frac{W_0^3}{8}\frac{1+2rp_c+2r^2p_c+r^3p_c^2}{(1-r^3p_c)(1-r^2p_c)(1-rp_c)}\\
            -\frac{3W_0^3}{8}\frac{1+2rp_c-2rp_c^2-2r^2p_c^2-2r^3p_c^2+2r^3p_c^3+r^4p_c^4}{(1-p_c)(1-rp_c)^2(1-r^2p_c)^2}\\
            +\frac{3W_0}{8}\frac{1-6rp_c^2+p_c(1+r)+rp_c^3(1+r)+r^2p_c^4}{(1-p_c)^2(1-rp_c)^3}
            -\frac{5p_c^2-4p_c+5}{8(1-p_c)^3}
        \end{pmatrix} \\
        &+\begin{pmatrix} T^3_{coll}p_c\frac{5p_c^2+4p_c+5}{(1-p_c)^3}
            + 3T^2_{coll}T_{succ}p_c\frac{1+p_c}{(1-p_c)^2}\\
            +3T_{coll}T^2_{succ}p_c\frac{1}{1-p_c}+T^3_{succ}\end{pmatrix}
            \\
        &+3A_1^2\begin{pmatrix} \frac{T_{coll}W_0^2}{4} \frac{p_c(1-r^2p_c^2)(1-2r^2p_c+r^2)+2rp_c(1-p_c)(1-r^2p_c)}
            {(1-p_c)(1-rp_c)^2(1-r^2p_c)^2}  \\
            -T_{coll}W_0\frac{p_c+rp_c+r^2p_c^2-3rp_c^2}{(1-p_c)^2(1-rp_c)^3}
            + \frac{T_{coll}}{2}\frac{3p_c^2-2p_c+2}{(1-p_c)^3}\\
            +\frac{T_{succ}W_0^2}{4}\frac{1+rp_c}{(1-rp_c)(1-r^2p_c)}
            -\frac{T_{succ}W_0}{2}\frac{1-rp_c^2}{(1-p_c)(1-rp_c)^2}\\
            +\frac{T_{succ}}{4}\frac{1+p_c}{(1-p_c)^2}
            \end{pmatrix}\\
        &+3A_1\begin{pmatrix} \frac{T_{coll}^2W_0}{2}\frac{rp_c-2rp_c^2-3rp_c^3+r^2p_c^2+p_c+p_c^2-3r^2p_c^3+4r^2p_c^4}
            {(1-p_c)^2(1-rp_c)^3} \\
            +T_{succ}T_{coll}W_0\frac{p_c(1+r-2rp_c)}{(1-p_c)(1-rp_c)^2}+\frac{T_{succ}^2W_0}{2}\frac{1}{(1-rp_c)}\\
            -\frac{T_{coll}^2}{2}\frac{8p_c^2-6p_c+4}{(1-p_c)^3}
            -T_{coll}T_{succ}\frac{2p_c}{(1-p_c)^2}-\frac{T_{succ}^2}{2}\frac{1}{(1-p_c)}
            \end{pmatrix}
    \end{split}
\end{equation}

\begin{biography}[{\includegraphics[width=1in,height=1.25in,clip,keepaspectratio]{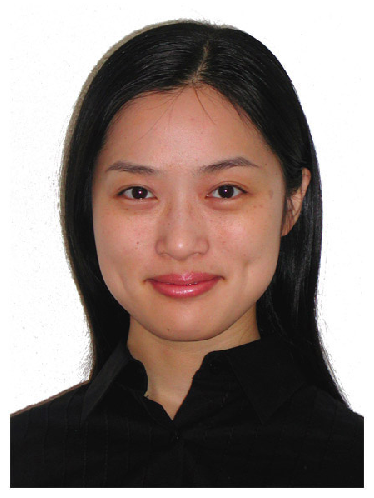}}]{Ying Jun (Angela) Zhang}
(S'00, M'05) received her PhD degree in Electrical and Electronic Engineering from the Hong Kong University of Science and Technology, Hong Kong in 2004. Since Jan. 2005, she has been with the Department of Information Engineering in The Chinese University of Hong Kong, where she is currently an assistant professor.

Dr. Zhang is on the Editorial Boards of IEEE Transactions of Wireless Communications and Willey Security and Communications Networks Journal. She has served as a TPC Co-Chair of Communication Theory Symposium of IEEE ICC 2009, Track Chair of ICCCN 2007, and Publicity Chair of IEEE MASS 2007. She has been serving as a Technical Program Committee Member for leading conferences including IEEE ICC, IEEE GLOBECOM, IEEE WCNC, IEEE ICCCAS, IWCMC, IEEE CCNC, IEEE ITW, IEEE MASS, MSN, ChinaCom, etc. Dr. Zhang is an IEEE Technical Activity Board GOLD Representative, 2008 IEEE GOLD Technical Conference Program Leader, IEEE Communication Society GOLD Coordinator, and a Member of IEEE Communication Society Member Relations Council (MRC).

Her research interests include wireless communications and mobile networks, adaptive resource allocation, optimization in wireless networks, wireless LAN/MAN, broadband OFDM and multicarrier techniques, MIMO signal processing.
As the only winner from Engineering Science, Dr. Zhang has won the Hong Kong Young Scientist Award 2006, conferred by the Hong Kong Institution of Science.
\end{biography}

\begin{biography}[{\includegraphics[width=1in,height=1.25in,clip,keepaspectratio]{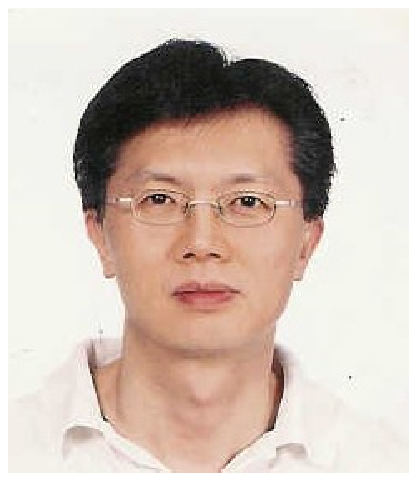}}]{Soung Chang Liew} (S'84-M'87-SM'92)
received his  S.B., S.M., E.E., and Ph.D. degrees from the Massachusetts Institute of Technology. From 1984 to 1988, he was at the MIT Laboratory for Information and Decision Systems, where he investigated Fiber-Optic Communications Networks. From March 1988 to July 1993, Soung was at Bellcore (now Telcordia), New Jersey, where he engaged in Broadband Network Research. Soung is currently Professor of the Department of Information Engineering,  the Chinese University of Hong Kong. He is Adjunct Professor at Southeast University, China.

Soung's current research interests include wireless networks, Internet protocols, multimedia communications, and packet switch design. Soung and his student won the best paper awards in the 1st IEEE International Conference on Mobile Ad-hoc and Sensor Systems (IEEE MASS 2004) the 4th IEEE International Workshop on Wireless Local Network (IEEE WLN 2004). Separately, TCP Veno, a version of TCP to improve its performance over wireless networks proposed by Soung and his student, has been incorporated into a recent release of Linux OS. In addition, Soung initiated and built the first inter-university ATM network testbed in Hong Kong in 1993.

Besides academic activities, Soung is also active in the industry. He co-founded two technology start-ups in Internet Software and has been serving as consultant to many companies and industrial organizations. He is currently consultant for the Hong Kong Applied Science and Technology Research Institute (ASTRI), providing technical advice as well as helping to formulate R\&D directions and strategies in the areas of Wireless Internetworking, Applications, and Services.

Soung is the holder of four U.S. patents and Fellow of IEE and HKIE. He is listed in Marquis Who's Who in Science and Engineering. He is the recipient of the first Vice-Chancellor Exemplary Teaching Award at the Chinese University of Hong Kong. Publications of Soung can be found in www.ie.cuhk.edu.hk/soung.
\end{biography}



\begin{thebibliography}{99}
\bibitem{IEEE:99}
IEEE standard for wireless LAN medium access control (MAC) and
physical layer (PHY) specifications, ISO/IEC 8802-11: 1999(E), 1999.

\bibitem{Bianchi:00}
G.~Bianchi, ``Performance analysis of the IEEE 802.11 distributed
coordination function'' \textit{IEEE J.~Select.~Areas Commun.,} vol.
18, no. 3, pp. 535--547, March 2000.

\bibitem{Bianchi:05}
G.~Bianchi and I.~Tinnirello, ``Remarks on IEEE 802.11 DCF
performance analysis,'' \textit{IEEE Commun. Lett.,} vol. 9, no. 8,
Aug. 2005.

\bibitem{Xiao:05}
Y. Xiao, ``Performance analysis of priority schemes for IEEE 802.11
and IEEE 802.11e wireless LANs,'' \textit{IEEE Trans. Wireless
Commun.,} vol. 4, no. 4, pp. 1506--1515, July 2005.


\bibitem{Kwak:05}
B.-J.~Kwak, N.-O.~Song, and L.~E.~Miller, ``Performance analysis of
exponential backoff,'' \textit{IEEE Trans. Network.,} vol. 13, no.
2, pp. 343--355, April 2005.

\bibitem{Sakurai:07}
T.~Sakurai and H.~L.~Vu, ``MAC access delay of IEEE 802.11 DCF,''
\textit{IEEE Trans. Wireless Commun.,} vol. 6, no. 5, pp.
1702--1710, May 2007.

\bibitem{Yang:03}
Y.~Yang and T.-S. Yum, ``Delay distributions of slotted ALOHA and CSMA," \textit{IEEE Trans. Commun.,} vol. 51, no. 11, pp. 1846-1857, Nov. 2003.

\bibitem{Zhai:05}
H.~Zhai, X.~Chen, Y.~Fang, ``How well can the IEEE 802.11 wireless
LAN support quality of service,'' \textit{IEEE Trans. Wireless
Commun.,} vol. 4, no. 6, pp. 3084--3094, Nov. 2005.

\bibitem{Duffy:05}
K. Duffy, D. Malone, and D. Leith, ``Modeling the 802.11 distributed coordination function in non-saturated conditions," \textit{IEEE Commun. Lett.}, vol. 9, no. 8, Aug. 2005.

\bibitem{Malone:07}
D.~Malone, K.~Duffy, and D.~Leith, ``Modeling the 802.11 distributed coordination function in non-saturated heterogeneous conditions," \textit{IEEE Trans. Network.,} vol. 15, no. 1, pp. 159-172, Feb. 2007.


\bibitem{Tickoo:04}
O.~Tickoo and B.~Sikdar, ``A queueing model for finite load IEEE
802.11 random access MAC,'' \textit{IEEE ICC04,} vol. 1, pp.
175--179, June 2004.

\bibitem{Naware:05}
V.~Naware, G.~Mergen, and L.~Tong, ``Stability and delay of
finite-user slotted ALOHA with multipacket reception,'' \textit{IEEE
Trans. Info. Theory,} vol. 51, no. 7, pp. 2636--2656, July 2005.

\bibitem{Zheng:06ICC}
P.~X.~Zheng, Y.~J.~Zhang, and S.~C.~Liew, ``Multipacket reception in
wireless local area networks,'' \textit{IEEE ICC06,} vol. 8, pp.
3670--3675, June 2006.

\bibitem{Zheng:06WLN}
P.~X.~Zheng, Y.~J.~Zhang, and S.~C.~Liew, ``Analysis of exponential
backoff with multipacket reception in wireless networks,''
\textit{IEEE Sixth International Workshop on Wireless Local Networks
(WLN),} pp. 852--862, Nov. 2006.


\bibitem{Verdu:98}
S.~Verdu, \textit{Multiuser detection,} Cambridge University Press,
Cambridge, UK, 1998.

\bibitem{Zhang}
Y.~J.~Zhang, P.~X.~Zheng, and S.~C.~ Liew, ``How does multiple-packet reception cabaibility scale the performance of wireless local area networks?'',
Technical resport, The Chinese University of Hong Kong.

\bibitem{Cali:00}
F.~Cali, M.~Conti, and E.~Gregori, ``Dynamic tuning of the IEEE
802.11 protocol to achieve a theoretical throughput limit,''
\textit{IEEE/ACM Trans. Network.,} vol. 8, no. 6, pp. 785--799, Dec.
2000.

\bibitem{Doshi:86}
B.~T.~Doshi, ``Queueing systems with vacations: A survey,'' Queueing
Systems, 1(1986)29--96

\bibitem{Kleinrock:75}
L.~Kleinrock, \textit{Queueing systems,} vol. 1: Theory, John Wiley
\& Sons Inc. 1975.

\bibitem{Soung}
S. C. Liew, Y. J. Zhang, and D. R. Chen, ``Bounded mean-delay throughput and non-starvation condition in
ALOHA network,'' to appear in \emph{IEEE Transactions on Networking}. Can be downloaded at http://arxiv.org/ftp/arxiv/papers/0801/0801.4054.pdf

\bibitem{Garetto:05}
M. Garetto and C.-F. Chiasserini, ``Performance analysis of the 802.11 distributed coordination function under sporadic traffic," \textit{Networking'05}, Waterloo, 2005.

\bibitem{Javidi:05}
T. Javidi, M. Liu and R. Vijayakumar, ``Saturation Rate in 802.11 Revisited", \emph{Annual Allerton Confercence on Communications, Control and Computing}, September 2005, Allerton, IL.

\bibitem{heavytail}
S. Asmussen, \emph{Applied probability and queues,} Springer, 2003.

\bibitem{Chan:04}
D. S. Chan and T. Berger, ''Performance and cross-layer design of CSMA for wireless networks with multipacket reception capability," \emph{The 38th Asilomar Conference on Signals, Systems and Computers}, vol. 2, pp. 1917-1921, Nov 2004.

\bibitem{Harsha:06}
S. Harsha, A. Kumar, and V. Sharma, ``An analytical model for the capacity estimation of combined VOIP and TCP file transfers over EDCA in an IEEE 802.11e WLAN," \emph{The 14th IEEE International Workshop on Quality of Service}, pp. 178-187, June 2006.

\end{thebibliography}
\end{document}